\documentclass[useAMS,usenatbib]{mn2e}

\usepackage{graphicx}
\usepackage{verbatim}   % for comments
\usepackage{longtable}
\usepackage{fixltx2e}   % to place figure and figure* in correct order
\newcommand{\bold}[1]{#1}               % for arXiv
\title[Long-term photometry and periods for 261 nearby pulsating M giants]{Long-term photometry and periods for 261 nearby pulsating M giants}
\author[V. Tabur, T.R. Bedding, L.L. Kiss, T.T. Moon, B. Szeidl and H. Kjeldsen]{V. Tabur$^{1}$\thanks{E-mail: tabur@physics.usyd.edu.au}
, T.R. Bedding$^1$, L.L. Kiss$^1$, T.T. Moon$^1$, B. Szeidl$^2$ and H. Kjeldsen$^3$
\\
$^{1}$Sydney Institute for Astronomy, School of Physics, A28, The University of Sydney, NSW 2006, Australia\\
$^{2}$Konkoly Observatory, Hungarian Academy of Sciences, H-1525 Budapest, P.O. Box 67, Hungary\\
$^{3}$Danish AsteroSeismology Centre (DASC), Department of Physics and Astronomy, University of Aarhus, DK-8000 Aarhus C, Denmark}

\begin{document}

\date{Accepted 2009 August 20. Received 2009 August 9; in original form 2009 May 24}

\pagerange{\pageref{firstpage}--\pageref{lastpage}} \pubyear{2009}

\maketitle

\label{firstpage}

\begin{abstract}

We present the results of a 5.5-year CCD photometric campaign that monitored 261 bright, southern, semi-regular variables with relatively precise Hipparcos parallaxes. The data are supplemented with independent photoelectric observations of 34 of the brightest stars, including 11 that were not part of the CCD survey, and a previously unpublished long time-series of VZ Cam. Pulsation periods and amplitudes are established for 247 of these stars, the majority of which have not been determined before. All M giants with sufficient observations for period determination are found to be variable, with 87\% of the sample (at S/N $\ge 7.5$) exhibiting multi-periodic behaviour. The period ratios of local SRVs are in excellent agreement with those in the Large Magellanic Cloud. Apparent $K$-band magnitudes are extracted from multiple NIR catalogues and analysed to determine the most reliable values. We review the effects of interstellar and circumstellar extinction and calculate absolute $K$-band magnitudes using revised Hipparcos parallaxes.
\end{abstract}


\section{Introduction}

Photometric surveys of M giants in the Large Magellanic Cloud (LMC), Small Magellanic Cloud (SMC), and Galactic Bulge have revealed a rich variety of pulsational behaviour. Pulsating M giants are potentially important as distance indicators since they obey period-luminosity (P-L) relations. The P-L relation for Miras has been known for decades \citep{b_egg,b_rob,b_gla}. More recent studies have shown that semi-regular variables (SRVs) also follow well-defined P-L relations. They may pulsate on the Mira sequence or in higher overtones, where they exhibit shorter periods and smaller amplitudes than the Miras \citep[see, for example,][]{b_woo96,b_bed98,b_woo99,b_kis03,b_kis04,b_ita04,b_gro04,b_wra,b_sos07}.

Meanwhile, studies of M giants in the local solar neighbourhood have been hampered by imprecise distances and a lack of detailed, long-term observations to enable the determination of characteristic periods. Using published periods for 24 galactic Miras and SRVs having relatively precise Hipparcos parallaxes, \citet{b_bed98} found two parallel sequences in the ($K$, $\log P$) plane, similar to those in the LMC. Further work by \citet{b_kna} and \citet{b_yes} increased the sample size and found evidence for at least two sequences, but the scatter due to imprecise parallaxes made it difficult to assign stars to distinct sequences. In the most extensive study to date, \citet{b_gla07} used revised Hipparcos parallaxes and literature periods for 64 red giants, confirming that local stars obey similar P-L sequences to those found in the Magellanic Clouds and Bulge. However, they concluded that details remain sketchy due to insufficient data, particularly for short-period stars.

Due to their complex multi-periodic behaviour, SRVs require monitoring over several years to determine their periods, particularly since many exhibit long secondary periods superposed with short-period, small-amplitude variations. Unfortunately, targets with good Hipparcos parallaxes are too bright for surveys such as ASAS \citep{b_poj97} and NSVS \citep{b_woz04}.

We have constructed a wide-field, automated survey telescope optimised for precise photometry of bright local SRVs, with the dual goals of determining periods for a sufficiently large sample to enable unambiguous identification of the local sequences, and to determine the effect of metallicity on the P-L zero-point. In this paper, we describe our sample selection (Sect. \ref{s_samp}), the selection of $K$-band magnitudes (Sect. \ref{s_k_mags}), CCD instrumentation (Sect. \ref{s_ccd_phot}), photometric pipeline and supplementary information (Sect. \ref{s_pipeline}), and PEP photometry (Sect. \ref{s_pep_phot}). The main result of the survey is presented in Sect. \ref{s_period_anal}, where we provide periods and $K$-band magnitudes, and discuss variability, multi-periodicity, and period ratios. We summarise our results in Section \ref{s_conclusion}. A detailed analysis of the results will be described in a follow-up paper, hereafter Paper II.

\section{Sample Selection}
\label{s_samp}
Our sample of nearby variable M giants was extracted from the Hipparcos catalogue \citep{b_per}, using the following criteria:
\begin{enumerate}
\item $\sigma_\pi / \pi \le 0.2$. Catalogue entries with negative parallaxes were ignored.
\item Coarse variability flag (H6) $\ge 2$, indicating that the Hipparcos mission had observed amplitudes $\ge 0.06$ mag. This was used to select true variables.
\item Spectral type (H76) = 'M*'.
\end{enumerate}
This yielded a complete sample of variable M giants to 20\% parallax precision. A subsequent check prior to publication revealed that two stars (HIP 56518 \& 85302) had been erroneously omitted from the sample.

The candidates were further filtered to select those likely to yield high-quality photometry:
\begin{enumerate}
\item Stars with declination $> 35$\degr{}  were removed to ensure that all candidates would transit at altitude $\ge 20$\degr{} at the observing site.
\item Stars fainter than $m_V = 8.0$ at maximum light were excluded, retaining only those bright enough to yield photometry at the 1\% level.
\item 33 stars listed in the General Catalogue of Variable Stars (GCVS) as having established periods $> 100$ d were removed in order to concentrate on those undergoing short-period, low-amplitude pulsation.
\end{enumerate}
This selection yielded 183 candidates. A further 67 stars with $\sigma_\pi/\pi < 0.26$ were added to increase the total number to 250, to provide sufficient targets for a 5-hour nightly observing run. Stars were evenly distributed in right ascension so that a similar number of fields would be observed each night, regardless of the season.

In hindsight, the declination-based selection criterion was too loose, resulting in the selection of a few stars with northern declinations that were only visible for short intervals each year and so have insufficient observations to determine their periods. In addition, one star near the south celestial pole (NSV 18694) was unobservable due to a physical obstruction near the observatory.

\begin{figure}
 \includegraphics[scale=0.7, angle=0]{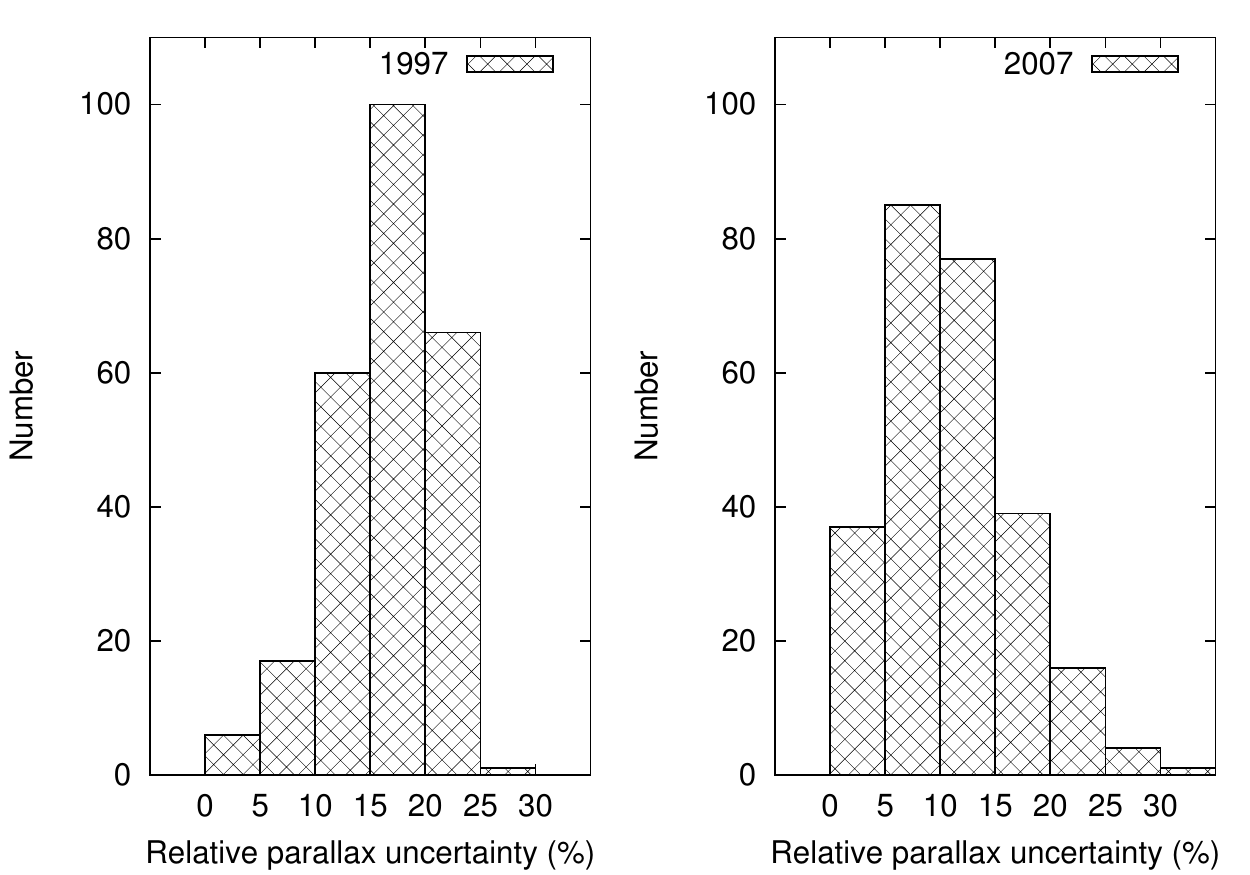}
 \caption{Distributions of relative uncertainty in parallax for the 250 M giant sample, using the original (1997) and revised (2007) Hipparcos catalogues, binned into 5\% intervals.}
 \label{fig001}
\end{figure}

Since our original sample selection, \citet{b_van07} has published revised Hipparcos parallaxes based on a re-reduction of the original data using an improved understanding of the rotation rates of the spacecraft \citep{b_van05}. The version published on VizieR on 15 September 2008,   which corrected an error that affected earlier versions of the catalogue, has been used for the analysis in this paper. The original and revised distributions of the relative parallax uncertainties of our sample are shown in Figure~\ref{fig001}. The revised distribution clearly shows an improvement for the majority of stars, with the median error reduced from 17.5\% to 10.6\%, consistent with the finding of \citet{b_gla07} that parallaxes for M giants were generally improved by a factor of 2. We have adopted the revised parallaxes for our analysis.

Our CCD data are supplemented by the following sources:
\begin{enumerate}
\item Twenty-three stars were observed concurrently by one of us (T.T.M.) using photoelectric photometry (PEP), as described in Section \ref{s_pep_phot}. An additional 11 stars were observed exclusively by PEP\bold{, including 4 stars of spectral type C, one S, and one K-giant ($\lambda$ Vel)}.
\item VZ Cam is a northern star with a relatively precise parallax measurement (4\%). We include a previously unpublished light curve obtained by B.Sz. at Konkoly Observatory.
\end{enumerate}
Note that OO Aps was initially observed by PEP only, but was later added to the CCD target list to check for the possible existence of short-period pulsations.

% Fig 3.1 [051/v3.eps]

\section{$K$-band magnitudes}
\label{s_k_mags}

In preparation for a Period-Luminosity (P-L) analysis (Paper II), we obtained apparent K magnitudes as follows.

\subsection{2MASS}

2MASS \citep{b_cut} has been the catalogue of choice for previous studies of pulsating red giants in the LMC, SMC, and Bulge. However, most of our M giants were saturated and have large uncertainties of 0.2--0.3 mag. Moreover, 2MASS magnitudes were derived from single-epoch measurements, leading to some variability-induced scatter, given that measured SRVs have $K$-band peak-to-peak amplitudes between 0.1 and 0.25 mag \citep{b_smi}. Thus, we only used 2MASS magnitudes when better values could not be found.

\subsection{The Gezari et al. Catalogue}

173 M giants in our sample appear in the Catalogue of Infrared Observations \citep{b_gez}. All magnitudes and colour magnitudes (flags M and C, respectively) observed at a wavelength of $2.2 \pm 0.05 \mu$m were combined to calculate a mean $m_K$ for each star, weighted equally because the catalogue did not provide measurement uncertainties.

Of the 80 stars with more than one catalogued magnitude, the mean scatter (excluding seven outliers) was $0.04 \pm 0.02$ mag. Seven stars had an abnormally large scatter. In two, this was caused by an apparent transcription error where the sign of the magnitude had been reversed ($\mu$ Gem and DM Eri). In other cases, a single discordant value, though not obviously wrong, was sufficiently different to yield a large scatter. As a precaution, mean magnitudes with $\sigma > 0.1$ mag were not used.

\subsection{DIRBE}

During its 10-month cryogenic mission, DIRBE collected 100--1900 measurements for each of nearly 12000 objects \citep{b_smi}. This permitted an accurate determination of mean $K$ magnitudes for short-period SRVs ($\sigma_{m_K} <$ 0.06 mag). Furthermore, DIRBE was less prone to saturation than 2MASS.

However, DIRBE had poor angular resolution due to its large beam width of 0.7\degr, which led to confusion between nearby NIR sources, particularly near the galactic plane ($|b| < 5$\degr). A search of the catalogue identified matches for 222 project stars, but only 166 were marked as unconfused sources. In 5 cases, quoted errors exceeded that of 2MASS (due to insufficient observations) and were thus excluded. Following \citet{b_whi}, we adopt a value of 630 Jy for $m_K=0$, which is in good agreement with \citet{b_bes}.

\begin{figure}
 \includegraphics[scale=0.7, angle=0]{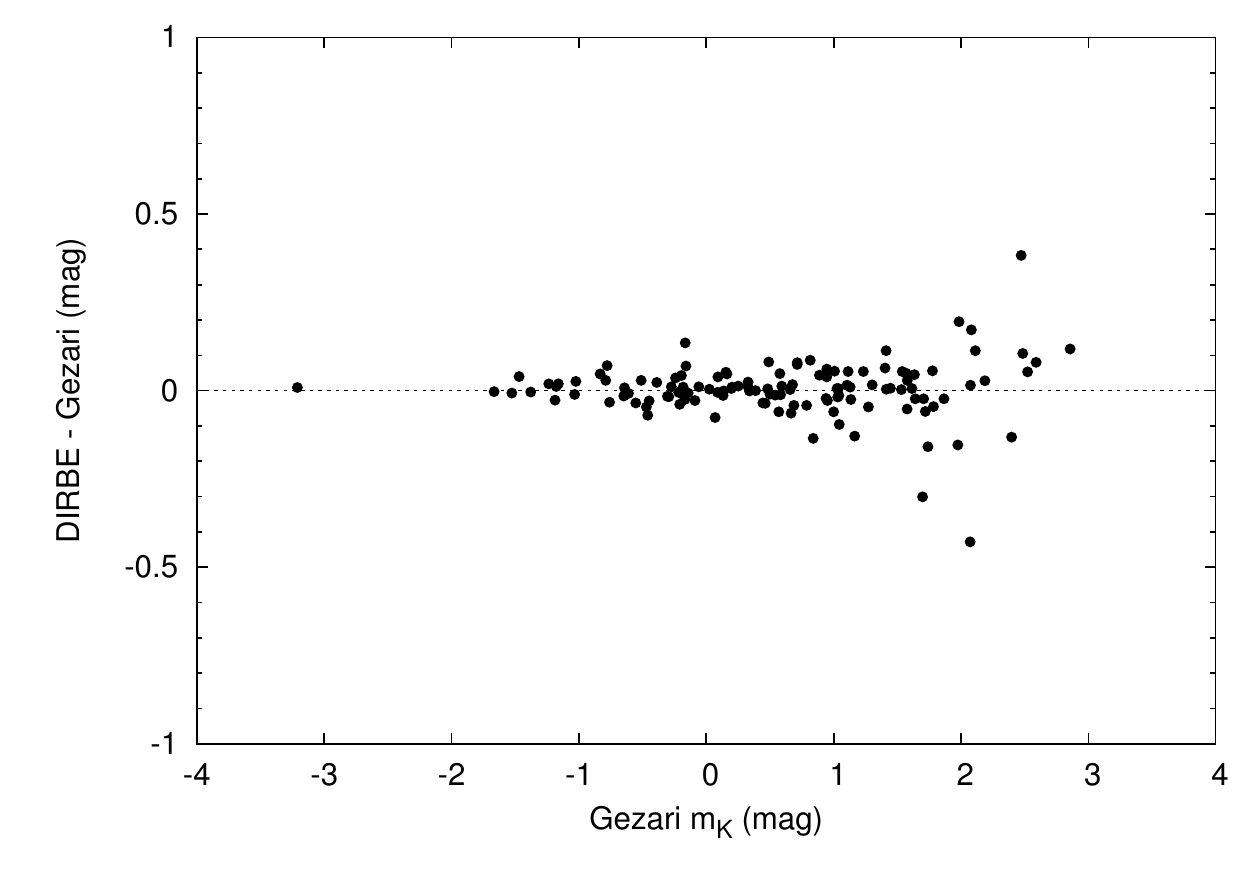}
 \caption{Apparent $K$-band magnitudes of  unconfused DIRBE sources compared to mean Gezari magnitudes, excluding stars from the latter with rms scatter $> 0.1$ mag, showing an excellent correlation.}
 \label{fig002}
\end{figure}

Figure~\ref{fig002} compares DIRBE $K$-band magnitudes with the mean $m_K$ values derived from the Gezari catalogue, excluding stars marked as confused in DIRBE and those stars extracted from Gezari with $\sigma > 0.1$ mag. The agreement is excellent, with a mean difference of $-0.004 \pm 0.081$ mag for the 125 stars on the plot, confirming that the calibration is appropriate. The small scatter is an indication of the consistency between the homogeneous DIRBE measurements and the mean values derived from multiple measurements taken from a set of inhomogeneous catalogues. Hence, it appears safe to use Gezari magnitudes when DIRBE values are unavailable or confused with other sources.

\begin{figure}
 \includegraphics[scale=0.7, angle=0]{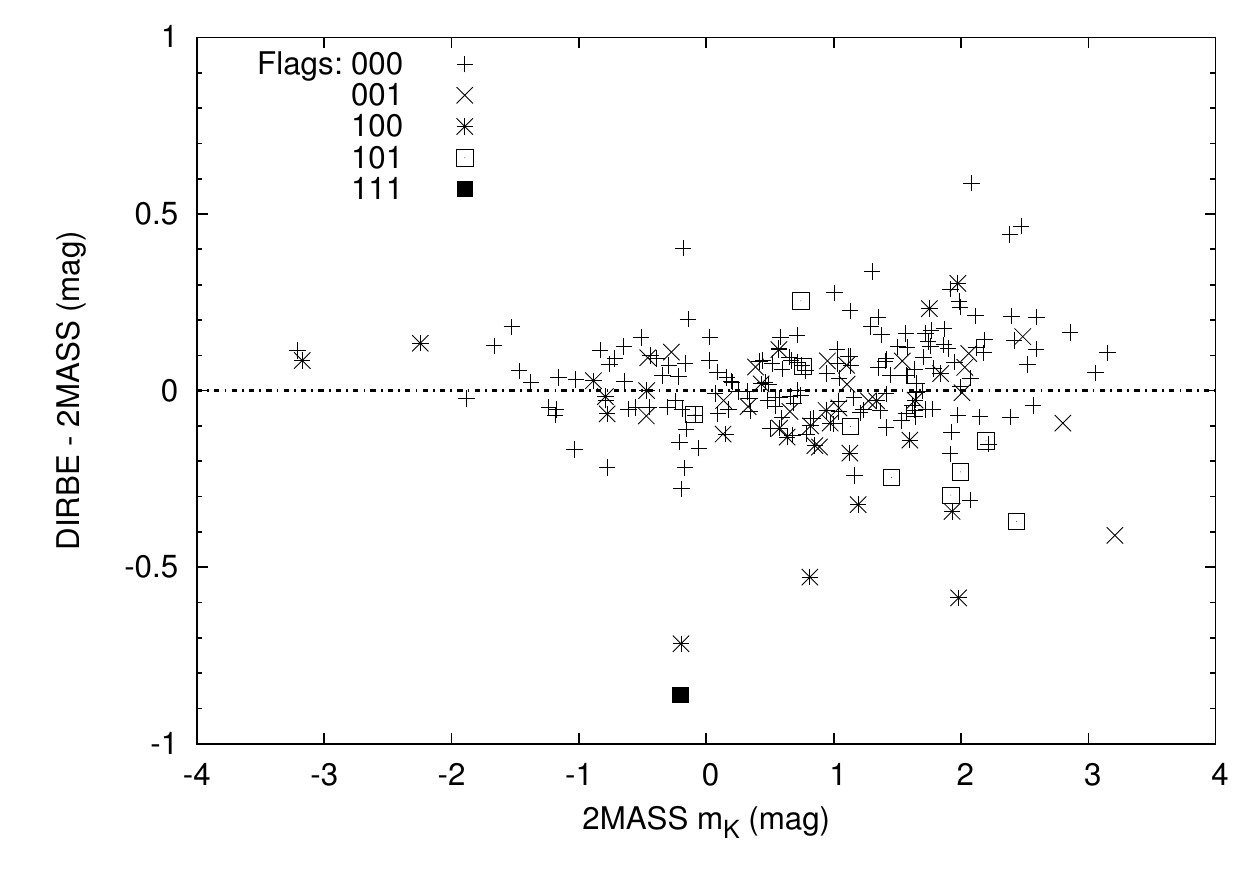}
 \caption{DIRBE vs. 2MASS $K$-band magnitudes using different symbols to denote various combinations of DIRBE confusion flags. The scatter is larger due to saturation in 2MASS.}
 \label{fig003}
\end{figure}

Figure~\ref{fig003} compares DIRBE magnitudes with 2MASS. The larger scatter than in Figure~\ref{fig002} is due to the saturation-induced errors in 2MASS magnitudes. The scatter for stars marked with confusion flag 3 (001) is no worse than that for unconfused sources (000), so we retained both. However, stars with flags 1 or 2 typically show an excess in the DIRBE measurement due to contamination with a nearby source, and have been excluded.

In summary, DIRBE $K$-band magnitudes were adopted except in cases where confusion flag 1 or 2 was set, or the DIRBE uncertainty exceeded that of 2MASS. In those cases, Gezari mean magnitudes were used when available \bold{(41 stars)}, otherwise the 2MASS magnitude was used \bold{(25 stars)}.

% Fig 3.9 [060/v1.png]

% Fig 3.10 [059/v1.png]

\subsection{Interstellar Extinction}

Since all stars in the sample are nearby, with the most distant being PS Hya (552 pc), we expect a relatively small amount of interstellar extinction, particularly in the $K$-band. Using a $(B-V)$ vs. $(V-I)$ colour-colour diagram, we found that the majority of stars are unreddened. The maximum colour excess of $E(B-V) \sim 0.2$ occurred for 7 stars with $V-I > 3.5$, implying a visual extinction of $A_V \sim 0.6$ mag. This was confirmed using two independent interstellar extinction models (\citealt{b_hak} and \citealt{b_dri03}), which indicated a maximum of $A_V \sim 0.5$ mag for all stars, with $A_V < 0.2$ mag for the majority. Calculating $K$-band extinction from $A_K / A_V = 0.112$ \citep{b_rie,b_sch}, we find an upper limit of $\sim 0.05$ mag with $A_K < 0.02$ mag for 90\% of the sample. Thus the correction is small, particularly since $M_K$ errors are dominated by much larger parallax uncertainties. Nevertheless, for the sake of completeness and to avoid systematic errors, corrections were applied using the model of \citet{b_dri03}, due to its superior resolution.

\subsection{Circumstellar Extinction}

Small-amplitude SRVs have smaller circumstellar envelopes than their high-amplitude, more evolved counterparts, the Miras. To investigate the possibility of reddening due to a dusty envelope, we tested for IR excesses at $12\mu m$ by plotting $V-I$ vs $V-[12]$ for the 231 stars in the sample having $12 \mu m$ flux densities listed in the DIRBE catalogue. Only a few stars showed a small excess. A catalogue of low-resolution spectra obtained by the \emph{Infrared Astronomical Satellite} (IRAS) confirmed that 8 stars had a small amount of circumstellar dust, while 15 showed silicate emission features. Using the relation of \citet[Equ 3]{b_kna} to estimate the $K$-band extinction caused by silicate dust, we found negligible amounts for the majority of stars ($A_K < 0.02$ mag), and thus we have ignored the effects of circumstellar extinction.

\section{CCD Observations}
\label{s_ccd_phot}

\subsection{Equipment}

\begin{figure}
 \includegraphics[scale=0.3, angle=0]{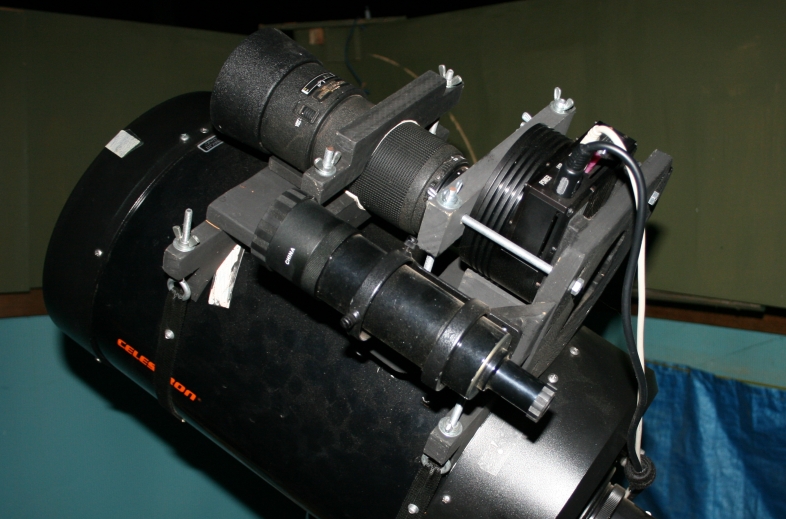}
 \caption{Survey equipment: Nikon 80--200-mm telephoto lens and SBIG ST8XE CCD detector attached to the telescope with a wooden cradle.}
 \label{fig004}
\end{figure}

An automated photometric telescope was constructed from relatively low-cost commercial components. A Nikon 80--200-mm $f$/2.8 ED telephoto lens ($D$=77 mm) coupled to an SBIG ST8-XE CCD detector with 9\,$\mu$m pixels was attached to a Losmandy G-11 German Equatorial Mount controlled by an Astrometric Instruments SkyWalker II stepper motor system (Figure~\ref{fig004}).

% Fig 3.2 [065/2147.png]

The telescope was housed in a roll-off roof observatory in a suburban backyard in Canberra, Australia, at latitude 35\degr S. Observations were gathered on nearly every clear night over a 5.5-year period using a fully automated software pipeline that controlled target selection and scheduling, telescope slewing, closed-loop pointing corrections, and camera operations (V. Tabur, PhD thesis, in prep.). In addition to the photometric survey described here, the observatory has also been used to discover a comet and two novae \citep{b_tab02,b_tab03a,b_tab03b}.

\subsection{Detector Calibration}

The detector bias, readout noise and gain were determined experimentally and found to be in close agreement with the manufacturer's stated values.

We used the procedure outlined by \citet{b_ber} to determine that our detector was linear below 47000 ADU to within 0.1\%. Stars with pixel values exceeding this limit (prior to calibration) were marked as saturated and were discarded by our photometric pipeline. Integrations shorter than 2 s were found to be non-linear due to a 0.02 s delay in closing the mechanical vane shutter. A shutter mapping correction was derived using the method of \citet{b_zis} but was found to be unnecessary in practice because integrations longer than 3 s were used, to reduce scintillation noise.

\subsection{Reducing Flux}

\begin{figure}
 \includegraphics[scale=1.0, angle=0]{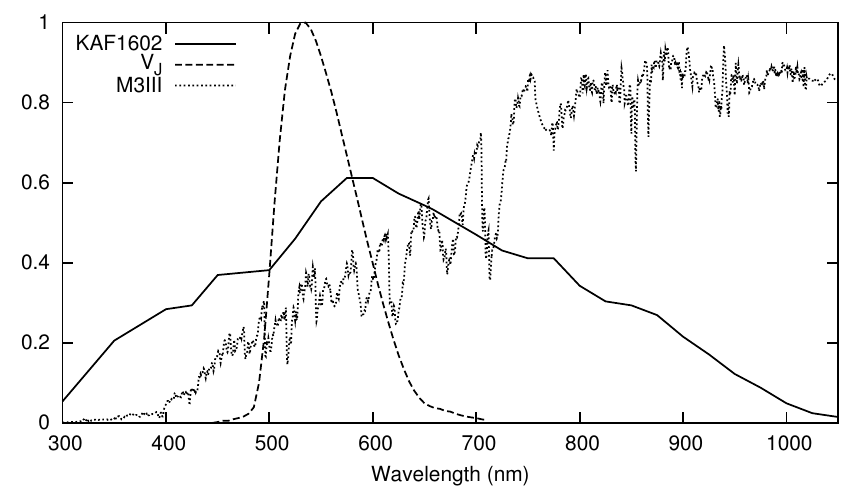}
 \caption{Quantum efficiency of the KAF1602 CCD sensor, relative transmission of a Johnson $V$-band filter (used for PEP), and the relative flux of a star of spectral type M3III, as a function of wavelength.}
 \label{fig004b}
\end{figure}

Three methods were employed to avoid detector saturation. By stopping-down the effective aperture of the lens to $50 \le D \le 64$ mm (during different phases of the survey) we reduced the number of photons reaching the detector. Secondly, short integration times of 3 s $\le t_{\rm int} \le 25$ s were employed, with the minimum practical limit set by scintillation noise and shutter effects. Broad-band photometric filters were not used because the SBIG CFW10 filter wheel does not provide sufficient back-focus when used in conjunction with a camera-lens adaptor. \bold{The unfiltered detector response is shown in Figure \ref{fig004b} and, for comparison, the relative transmission of a Johnson $V$-band filter, and the spectral energy distribution of a star of type M3III \citep{b_pic98}.}

The third way to reduce saturation was to distribute the flux over a large number of pixels. Stellar images were broadened by defocusing the optics, which also helped to improve precision by reducing sensitivity to the gate structure of the front-illuminated CCD. However, defocusing produced a complex PSF due to the multi-element optical design, and caused some blending in crowded fields, making it difficult to find isolated comparison stars and requiring more complicated analysis.

% Fig [097/x.plt]

\subsection{Automation}

Once initialised each night, the telescope operated completely autonomously, allowing a very efficient survey. The observing cycle consisted of the following tasks:
\begin{enumerate}
\item Target selection: In general, fields were ordered by increasing R.A. so that those about to set were observed first. Allowances were made for exclusion zones to prevent frequent meridian flips when targets were in the overlap zone (visible from both sides of the mount), and to avoid obstacles. Short slews were weighted more highly to increase the duty-cycle.
\item Open-loop slewing: A telescope pointing-model \citep{b_wal} was used to compensate for mechanical imperfections such as polar axis misalignment, non-orthogonality of the axes, and declination axis flexure. Pointing errors were $<$ 1 arcmin when slewing to the same side of the mount, and $<$ 3 arcmin all-sky. Slews were always completed with a final movement toward the W and S to remove backlash from the gear train.
\item Closed-loop pointing correction: Residual pointing errors were corrected using a micro-slew, with the necessary offsets determined from a 2-s, low-resolution integration. The plate solution was calculated using the \emph{Optimistic Pattern Matching} algorithm \citep{b_tab}, which was specifically designed to handle wide fields of view containing thousands of stars. The additional integration helped to flush the detector of any potential Residual Bulk Image caused by near saturation of the detector \citep{b_res}, while the final pointing correction reduced the effects of residual flat-fielding errors \citep{b_eve}.
\end{enumerate}

\subsection{Survey Phases}

Our survey consisted of three phases (Table \ref{T1}). Initially, a focal-length of $f=142$\,mm was employed with integrations ranging from 1 s $\le t_{\rm int} \le$ 25 s, and with the image slightly defocused. The detector was operated at full resolution with sub-frames read out from the central 1/3 of the CCD. The brightest targets required short exposures of $\sim$ 1--2 s, which led to a high level of scintillation noise. Despite the short integrations, some data were saturated and had to be discarded.

Phases 2 and 3 consisted of greater levels of defocus, plus an increased focal length to offset the blending of stellar sources. The sub-frame size was enlarged to increase the probability of finding bright comparison stars, which were required for the ensemble photometry.

Phase 3 used strongly defocused images yielding Full-Width at Half-Maximum (FWHM) values up to 25 pixels. Despite the difficulties this caused during analysis, requiring blended stellar sources to be deconvolved, this approach was adopted because it permitted the use of 25-s integrations for all but the very brightest targets, thus reducing scintillation noise to $\sim$ 0.5\%. Although this could have been reduced further with longer integrations, the limit was set to allow $\sim$125 fields (one visible hemisphere) to be visited within a 5-hour observing window.

Multiple consecutive integrations were obtained for each target field to measure internal precision and reduce scintillation noise. A series of 3, 5, 7, or 9 images were acquired for each field utilising 25, 12, 6, or 3 s integrations, respectively. The images were analysed separately and instrumental magnitudes combined to yield mean nightly values for all targets and their comparison stars.

Even the Phase 3 regime was insufficient to prevent the two brightest stars, $\gamma$ Cru and $\beta$ Gru, from saturating the detector when observed near the meridian, although some useful images were obtained at higher airmass. Their light curves were supplemented with PEP data (Section \ref{s_pep_phot}).

\begin{table}
  \caption{Survey Phases. Observations were concluded on 2008 Nov 5.}
  \label{T1}
  \setlength{\tabcolsep}{2mm}
{\scriptsize
  \begin{tabular}{@{}lllrllll}
  \hline
 Ph & Start Date  & JD      & Span  & Scale          & FOV            & FWHM\\
    &             &         & (days)& ($\arcsec$/px) & (deg)          & (px)\\
  \hline
  1 & 2003 May 28 & 2452787 & 731   & 13             & 1.9$\times$1.2 & 5\\
  2 & 2005 May 28 & 2453518 & 15    & 10             & 2.2$\times$1.5 & 10\\
  3 & 2005 Jun 12 & 2453533 & 1243  & 10             & 2.9$\times$2.0 & 18--25\\
  \hline
\end{tabular}
}  % scriptsize
\end{table}

\section{CCD Photometry}
\label{s_pipeline}

\subsection{Flat-Fielding}

Reliable flat-field images are notoriously difficult to produce for wide fields. Twilight flats have been shown to contain gradients of 1\% per degree when obtained too far from the null-point, where sky-illumination gradients are minimised \citep{b_chr}, while blank sky flats are non-uniform over $0.5\degr$ scales due to natural altitude-dependent gradients \citep{b_zho}.

We used a combination of approaches to solve these problems. Firstly, high S/N dome flats were acquired monthly to model the high-frequency, inter-pixel variations in detector response. Illumination gradients of $\sim$ 1\% were present, despite placing a two-stage diffusing screen near the entrance pupil of the optics \citep{b_zho}. To remove the gradients, each dome flat was corrected to match the low-frequency variations present in a master sky-flat (created by median combining 59 images obtained on a moonless night with the telescope pointing at the zenith), modelled using a 3rd-order, 2-dimensional polynomial. Finally, since the uniformity of the master sky-flat was not assured, we used the fact that an imperfect flat-field calibration will lead to spatially dependent photometric errors to determine an illumination correction. Since we observed each field hundreds of times, on both the east and west sides of the mount, the $180\degr$ rotation of the detector that occurs after a meridian-flip was used to model the change in mean magnitude of each comparison star as a function of its spatial location on the detector, analogous to the dithering of stars across an image to build a Photometric Super Flat \citep[see][]{b_man,b_boy}. In this way, we measured gradients in the master sky-flat of up to $\pm$ 6 mmag, which we corrected in the photometric pipeline.

\subsection{Atmospheric extinction}

To a second-order approximation, the true magnitude of a star, $V_0$, measured outside of the Earth's atmosphere, is related to the instrumental magnitude, $V_{\rm inst}$, by
\begin{equation}\label{eq:13}
V_0 = V_{\rm inst} - k' X - k'' X C,
\end{equation}
where $X$ is the airmass, $C$ is the colour-index of the star and $k'$ and $k''$ are the first- and second-order extinction coefficients. By using long-term observations of the comparison stars gathered over a wide range of airmasses, we determined mean values for $k'$ and $k''$ to be 0.16 mag airmass$^{-1}$ and 0.024 mag per airmass per unit colour, respectively. We applied the correction in the pipeline. A $V-I$ colour index was used for the colour term, to match the wide spectral response of the unfiltered detector, which spans 450 nm at FWHM and extends to the $I$ band where M giants radiate much of their energy.

\subsection{Photometric precision}

Magnitudes of targets and comparison stars were determined using aperture photometry. Faint, defocused stars in the sky-annulus of some comparison stars, particularly at low galactic latitudes, caused a systematic increase in sky-background measurements during bright moonlit periods, when they became difficult to separate from the real sky-background using a sigma-clipping routine. The problem was solved by mapping the positions of these stars and excluding their pixels from the sky annulus.

\begin{figure}
 \includegraphics[scale=0.7, angle=0]{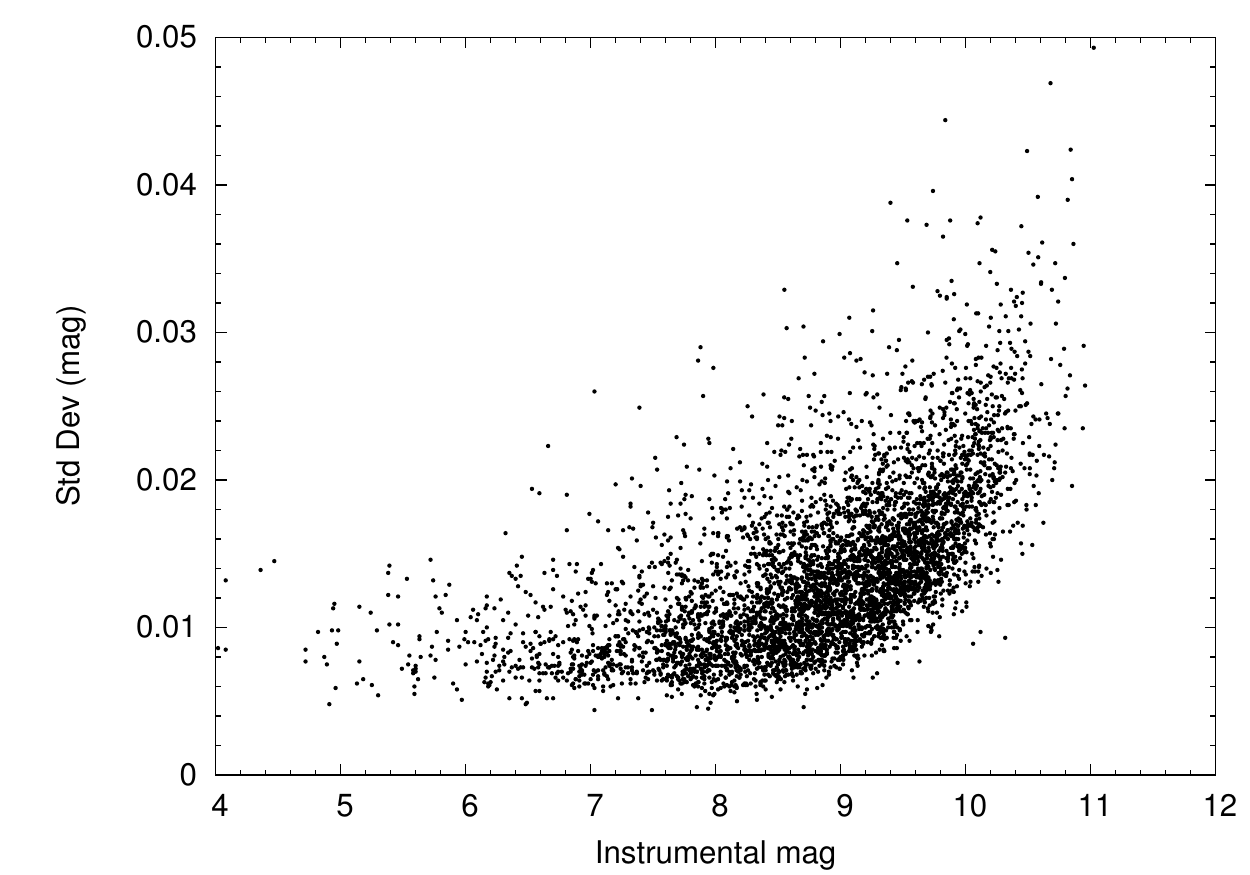}
 \caption{Individual comparison star scatter after correction of sky background estimates in crowded fields. Each measurement is the mean of three 25 s integrations with airmass $< 2.0$ (4979 stars).}
 \label{fig005}
\end{figure}

The photometric precision, measured as the rms scatter, is shown in Figure~\ref{fig005} for 4979 comparison stars. Our target stars have instrumental magnitudes brighter than 9, so they all fall in the more precise tail of the distribution. The best photometric precision achieved is limited to $\sim$ 6 mmag, even for the very brightest stars, which indicates the scintillation noise due to the small aperture and short integration times. \citet{b_you} described the well-known relation for atmospheric scintillation, $S$, the ratio of rms intensity fluctuation to mean intensity, as
\begin{equation}\label{eq:11}
S =  0.09  d^{-2/3} X^{3/2} \exp(-h/8000) / \sqrt{2 t_{\rm int}},
\end{equation}
where $d$ is the telescope aperture in cm, $X$ is the airmass, $h$ is altitude of the observing site in metres, and $t_{\rm int}$ is the integration time in seconds. Substituting values of $d$ = 6.4 cm, $h$ = 641 m, and $t_{\rm int}$ = 3 $\times$ 25 s for observations conducted during Phase 3, and assuming a mean airmass of $X$=1.5, we estimate the contribution due to scintillation to be $\sim 7$ mmag, which agrees well with the lower limit on the measured precision.

\subsection{Ensemble photometry}

Up to 50 of the brightest comparison stars in each field were combined to form an ensemble. We assumed constancy of the ensemble average, that is,
\begin{equation}\label{eq:12}
\frac{\sum{m_i w_i}}{\sum{w_i}} + ZP = 0,
\end{equation}
where $m_i$ are the instrumental magnitudes of the comparison stars, $w_i$ are their weights, and $ZP$ is the zero-point of each image in the time-series. The ensemble was used to determine iteratively the $ZP$ for each image \citep{b_bal}, permitting the scatter of each comparison star to be determined, unstable stars to be excluded, and weights to be assigned to the remainder. Following \citet{b_gil93}, weights were assigned using the reciprocal of their variance, normalised to a mean weight of 1, and clipped to a maximum value of 2 to prevent bright stars with low-amplitude variability from perturbing the solution.

Aperture photometry in crowded fields was compromised by the strong defocus, which caused some blending of stellar sources. Comparison stars were generally unaffected, since they were manually selected to avoid nearby companions. In cases where some contamination was unavoidable, the stars exhibiting the largest errors were excluded from the ensemble.

\begin{figure}
 \includegraphics[scale=0.7, angle=0]{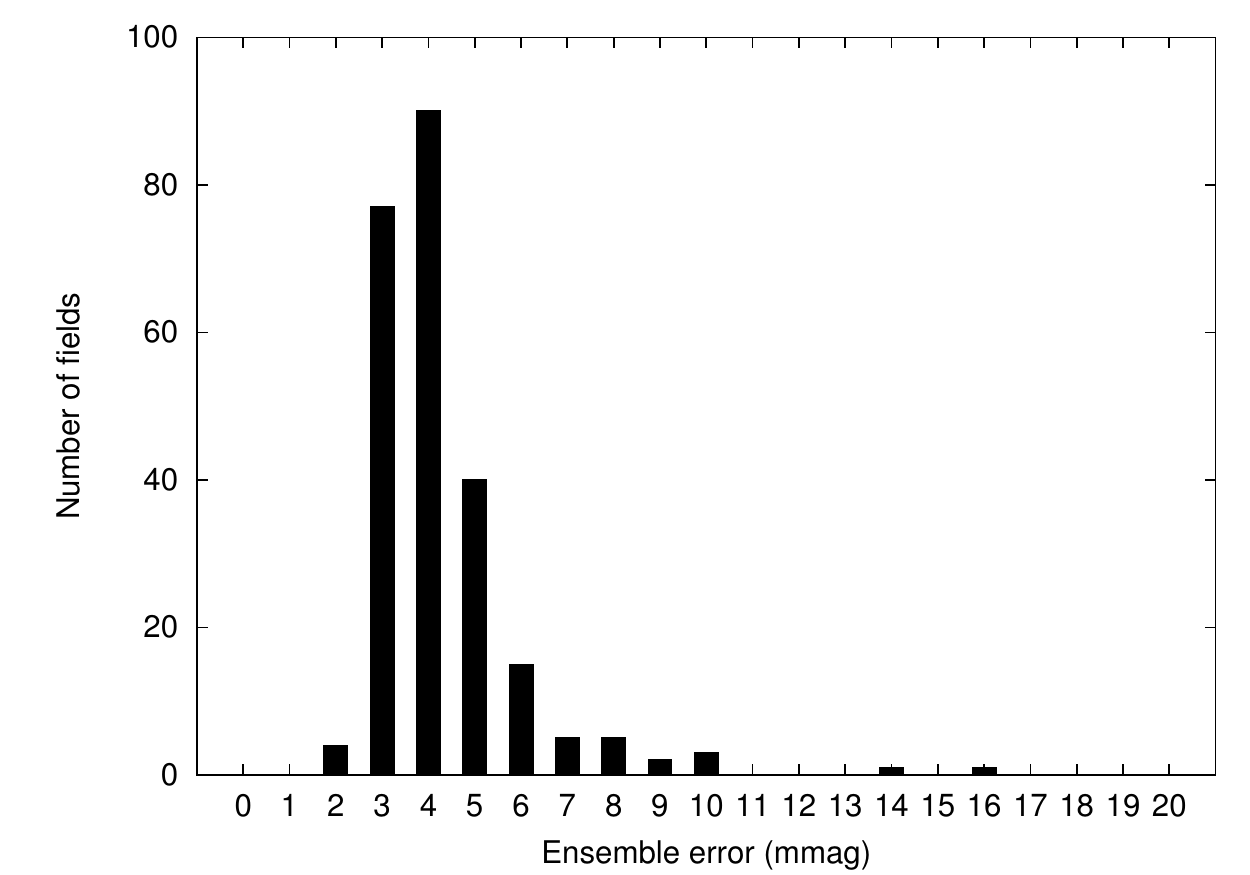}
 \caption{Distribution of mean ensemble uncertainty for 243 survey fields using 1 milli-mag bins.}
 \label{fig006}
\end{figure}

Figure~\ref{fig006} shows the distribution of the uncertainties for our ensemble of comparison stars, which were mostly about 3--5 mmag. Since the M giants were typically the brightest stars in each field, their internal errors were small, resulting in a total uncertainty for their differential magnitude determination of 7--8 mmag.

% Fig 2.22 [037/x.eps]

% Fig 2.27 [049/v1.eps]

\subsection{Optimal Image Subtraction}
\label{s_ois}

% K125 =  40144
% K244 =  44281
% K174 =  46027
% K195 =  64117
% K211 =  73764
% K62  =  80047
% K122 =  83258
% K158 =  87390
% K31  = 101810

Nine M giant targets\footnote{HIP 40144, 44281, 46027, 64117, 73764, 80047, 83258, 87390, 101810} were blended with nearby stars and required a Difference Image Analysis technique to subtract the flux contributed by their companions, allowing the variable sources to be measured free of contamination. Following \citet{b_ala}, we implemented an Optimal Image Subtraction (OIS) pipeline. High S/N images taken under the best seeing conditions were selected as reference images. Due to a small, spatially dependent asymmetry in the defocused PSF, two reference frames were selected for each field, corresponding to pre- and post-meridian-flip orientations. A high-order 2D polynomial was used to register each image with its reference to cater for the presence of optical distortions caused by the fast ($f$/2.8) optics. Well-focused images yielded residuals of $\sim$0.05 pix, but the Phase 3 images had larger residuals ($\sim$0.15 pix), due to the difficulty in determining accurate centroids for the defocused PSFs.

Since our survey fields were not crowded, we followed \citet{b_koc} by only using the pixels associated with the 50 brightest stars, to prevent the Poisson noise of the blank sky background from contributing to the kernel solution. Since the tuning of OIS parameters is not straightforward, we adopted the method and recommendations of \citet{b_isr} to determine optimal values.

% Fig 2.33 [056/x.eps] (and possibly 057)
\begin{figure*}
 \includegraphics[scale=1.0, angle=0]{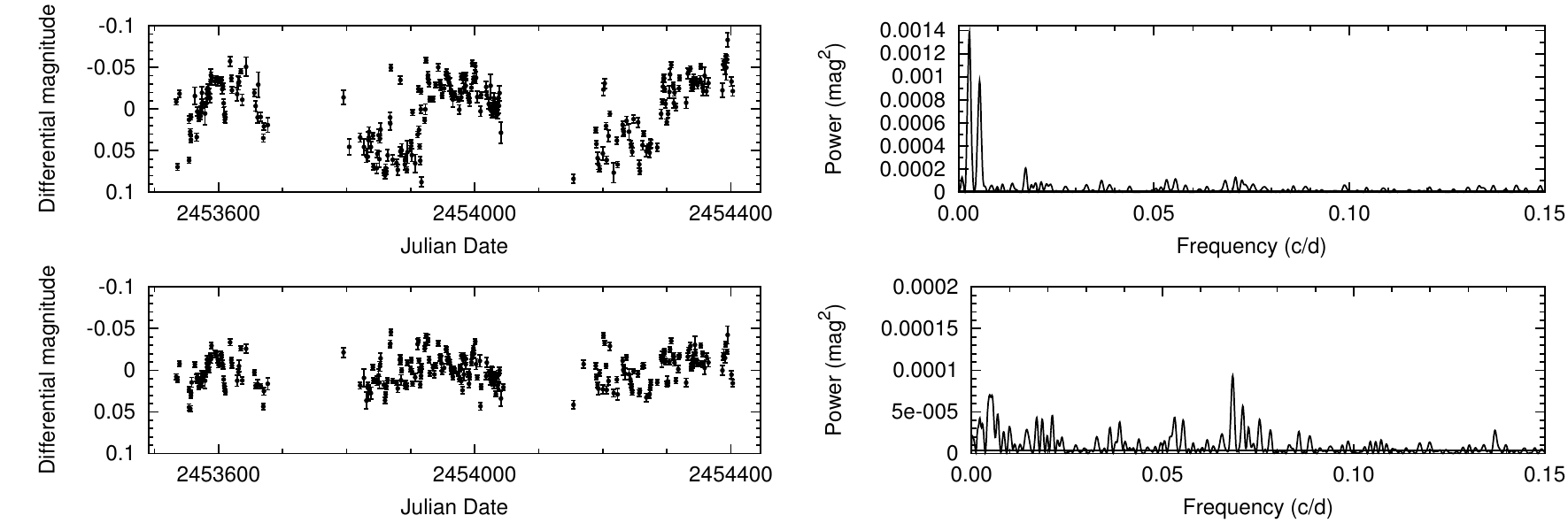}
 \caption{Light curve and power spectrum of NSV 20880 (HIP 83258), which was blended with a bright companion, before (top) and after processing with OIS (bottom), resulting in the removal of the 1- and 0.5 yr peaks which dominate the original power spectrum. Note the change in scale of the corrected power spectrum.}
 \label{fig007}
\end{figure*}

\begin{figure}
 \includegraphics[scale=1.0, angle=0]{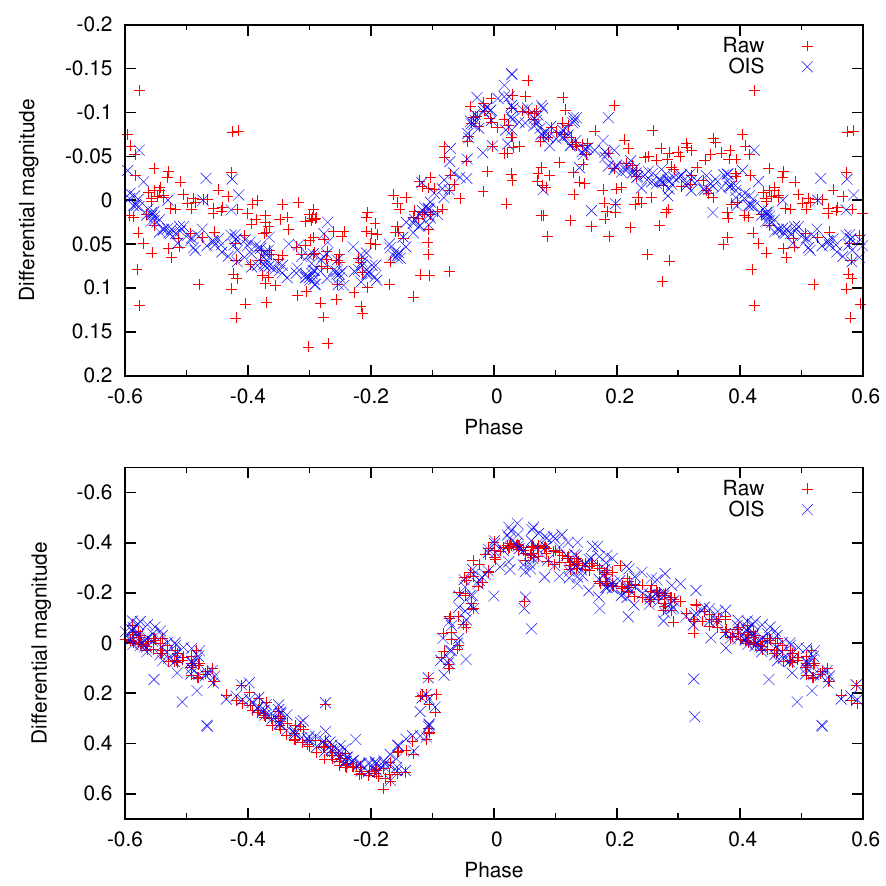}
 \caption{Phased light curves for the Cepheids AT Pup (top) and RS Pup (bottom), showing raw differential magnitudes (red pluses) and those following the application of OIS (blue crosses). AT Pup is blended with a brighter companion, leading to significant seeing-dependent aperture contamination, while RS Pup does not have a bright companion.}
 \label{fig008}
\end{figure}

We found that our OIS pipeline greatly improved the photometry in crowded fields. The noise in the power spectra of light curves produced using OIS were reduced, as were the strengths of spurious peaks at periods of 1 and 1/2 yr (Figure~\ref{fig007}). We used two Cepheids (AT Pup and RS Pup), located in the field of V427 Pup, as test cases. Both have well-known periods and AT Pup is strongly blended with a brighter companion. Figure~\ref{fig008} shows their phased light curves, and demonstrates the dramatic improvement in photometric precision when using OIS on blended objects (upper panel). \bold{In some cases OIS did not produce optimal results, such as the slight increase in scatter for RS Pup, which we attribute to sampling issues and numerical instabilities. We restricted its use to 9 blended targets.}

% 092/x.eps

\section{Photoelectric Photometry}
\label{s_pep_phot}

Independent PEP data for 23 of the project stars were obtained by one of us (T.T.M.) as part of a survey of southern SRVs \citep{b_moo}. The PEP observations span a similar temporal range to the CCD survey, but are generally sparser due to the manual operation of the equipment. Each measurement is the mean of 5 consecutive 10-second integrations using $V$ and $B$ broadband filters, obtained in the sequence $V_{\rm star}$, $B_{\rm star}$, $B_{\rm sky}$, $V_{\rm sky}$, and bracketed by observations of two comparison stars. Measurements were corrected for extinction and transformed to standard $V$ magnitudes and $B-V$ colour indices using data from the \emph{General Catalogue of Photometric Data} (\citealt{b_mer}).

\subsection{Transformation to $V_J$}

In order to combine CCD and PEP data, the unfiltered CCD observations of the 23 stars were transformed to Johnson $V$ magnitudes using numerous comparison stars in each field. Since some comparison stars were likely to be variable, ensembles were used to identify stable stars and their mean instrumental magnitudes. In the absence of colour information in the instrumental system, the $(V-I)_C$ colour index of each comparison star was used to fit the relation
\begin{equation}\label{eq:1}
	V_J - V_{\rm ins} = c_{0,f}  + c_1 (V-I) + c_2 (V-I)^2.
\end{equation}
Here, $V_J$ is the Johnson $V$-band magnitude, $V_{\rm ins}$ is the mean instrumental magnitude, $c_{0,f}$ is a zero point for each field, and $c_1$ and $c_2$ are the colour terms. Separate zero-points were required for each field because the ensemble used to derive the mean instrumental magnitudes consisted of an inhomogeneous set of stars and weights.

\begin{figure}
 \includegraphics[scale=0.7, angle=0]{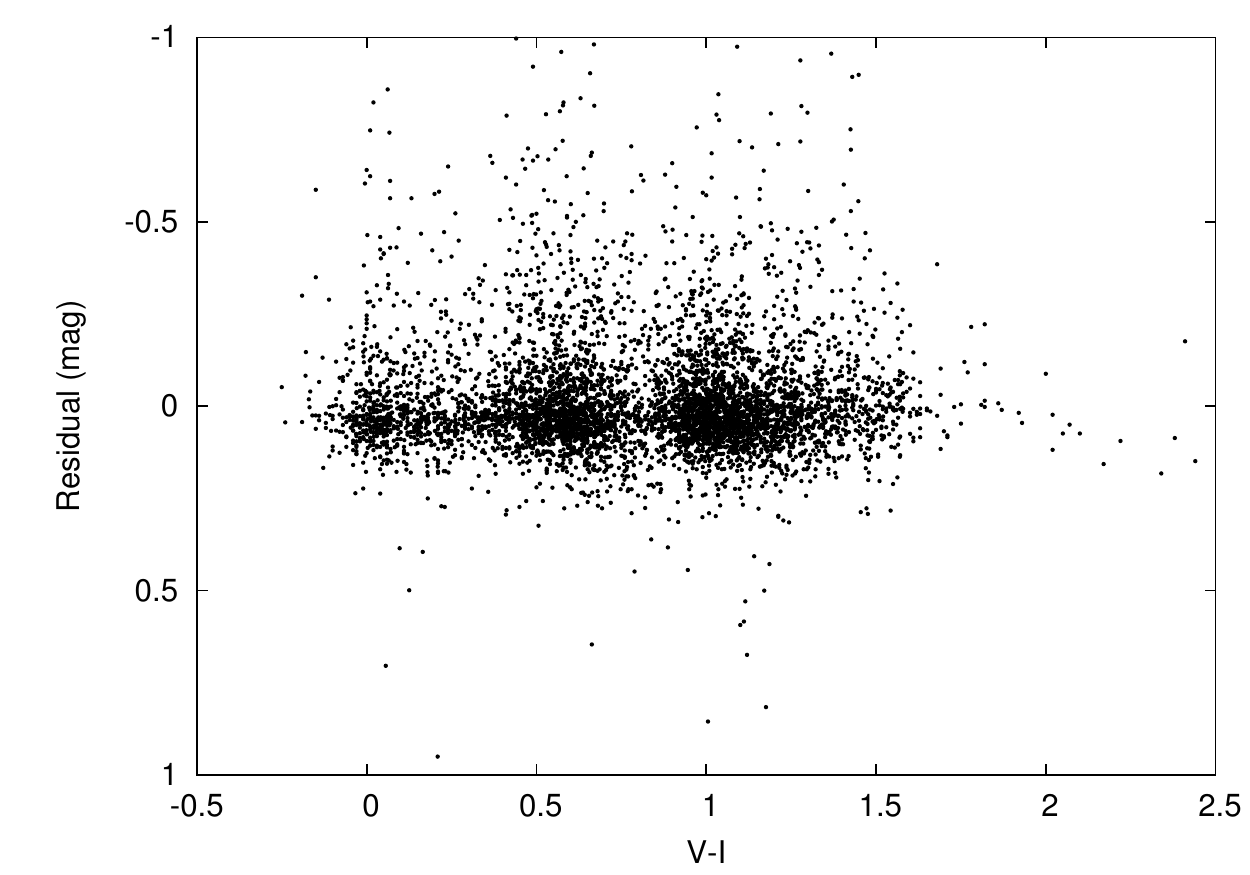}
 \caption{The difference between catalogued $V_J$ and the transformed values as a function of $V-I$ colour index. Stars with an excess are involved with a close companion that increased the measured flux within the photometric aperture, resulting in the asymmetrical distribution.}
 \label{fig009}
\end{figure}

After fitting, some stars had an excessively large difference between the catalogued $V_J$ and the transformed value, caused by contamination of their photometric aperture by a nearby companion, resulting in the asymmetry in Figure~\ref{fig009}. Outliers were removed using 3 iterations of a sigma clipping routine where stars $> 2.5\sigma$ from the median residual were removed. 4562 stars were retained in the final list of well-fitting, uncontaminated comparison stars, resulting in a scatter of 0.06 mag.

% Fig 3.3 [062/v2/v1.eps]

\begin{figure}
 \includegraphics[scale=0.7, angle=0]{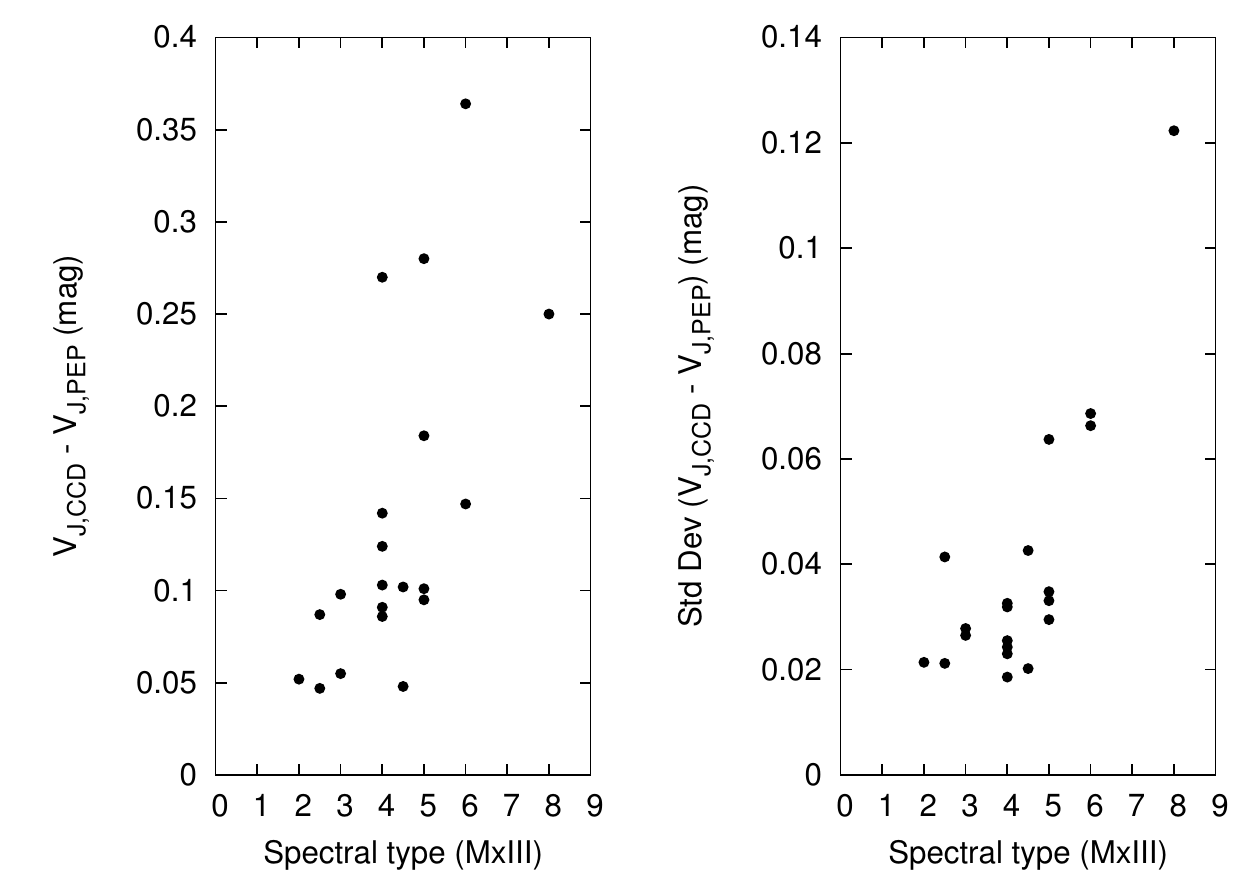}
 \caption{ Mean difference between M-giant $V_J$ magnitudes in the CCD and PEP systems as a function of Morgan-Keenan spectral type (left), and the resulting $1\sigma$ scatter in the differences after applying a zero-point correction as described in the text (right). Late-type stars exhibit greater errors due to increasing differences in effective passband.}
 \label{fig010}
\end{figure}

Although the fit was reasonable for the comparison stars, the cooler M giants exhibited larger residuals with systematic differences between the two systems (Figure~\ref{fig010}, left panel). We attribute these to the lack of instrumental colour information for use in the transformation, together with the use of an unfiltered detector that was sensitive to a broad range of wavelengths, and extrapolating the relation to cool temperatures beyond the range of comparison star colour indices $(V-I > 1.7)$. The left panel shows an approximately linear systematic difference between $V_J$ in the CCD and PEP systems for the M giants, as a function of spectral type. The filtered PEP data are approximately 0.05 mag brighter than the unfiltered CCD for M2III stars, and the difference increases to 0.37 mag for later-types. A hint of the same trend is visible in Figure~\ref{fig009} for $2 < V-I < 2.5$, corresponding to spectral types up to M2III, though there are too few stars to be certain, since most M giant comparison stars were variable and were thus excluded from the fit.

To adjust for the difference, a zero-point correction was added to align the CCD and PEP light curves. The correction was determined by using the CCD observations that bracketed each PEP observation by $\pm 2$ days to interpolate the CCD magnitude to the same date as the PEP observation. The mean difference between the CCD and PEP magnitudes defined the offset.

Figure~\ref{fig010} (right panel) shows the rms uncertainty of $(V_{J,\rm CCD} - V_{J,\rm PEP})$ residuals after applying the zero-point correction, as a function of spectral type. Most stars have a small scatter of $\sim$ 0.02 -- 0.04 mag, somewhat larger than the uncertainty of the PEP measurements, which were generally better than 0.01 mag. We attribute the differences to errors in the transformation. However, as expected, the coolest stars exhibit a larger scatter, caused by a systematic increase in passband difference due to the transmission of longer wavelengths on the red side of the $V$-band cut-off, to which our unfiltered CCD is sensitive.

\begin{figure*}
 \includegraphics[scale=1.0, angle=0]{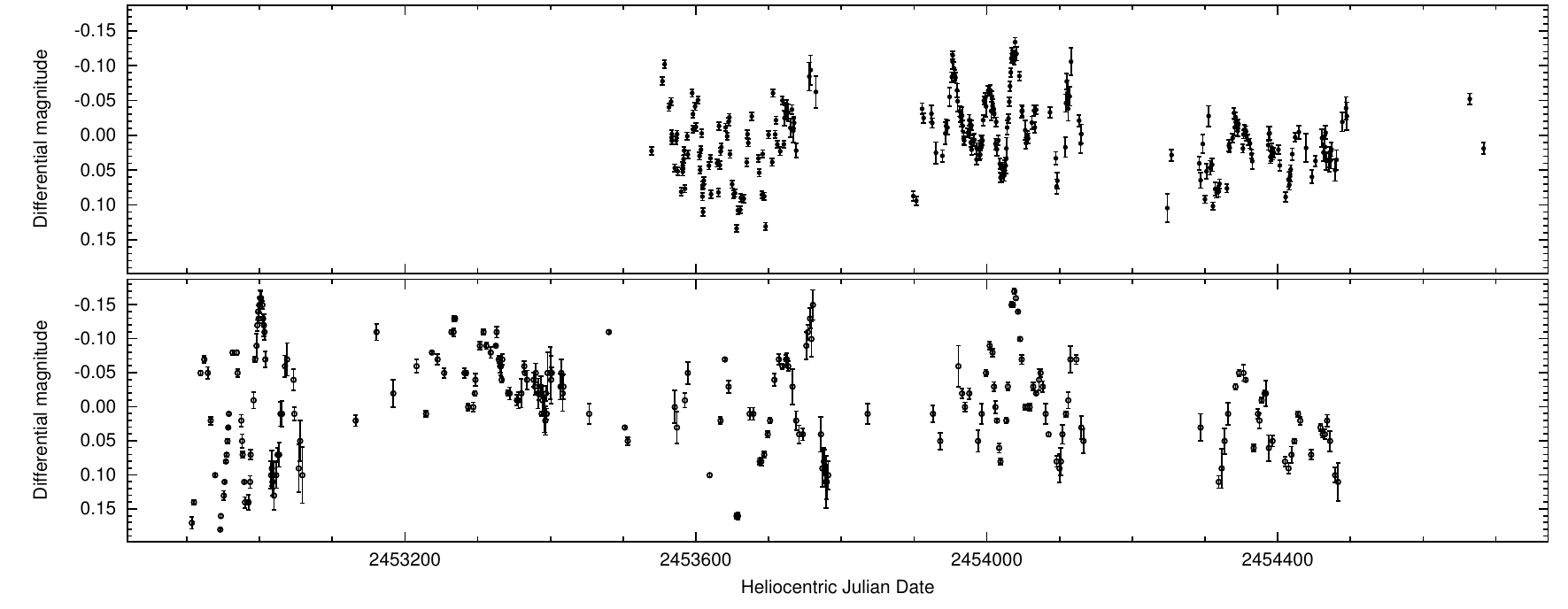}
 \caption{CCD (top panel) and PEP light curves of $\beta$ Gru. The agreement is good, with the PEP observations exhibiting a slightly larger photometric amplitude.}
 \label{fig011}
\end{figure*}

The PEP and CCD light curves exhibit the same general features, although the former exhibit greater $V$-band amplitudes (Figure~\ref{fig011}). The most extreme example occurs for the M8III variable V744 Cen, which has a 36\% greater amplitude in the PEP system. Fortunately, the coolest stars exhibit long-period, large-amplitude variations and so their periods are well determined using CCD data alone (Table \ref{T2}).

\begin{table}
  \caption{Late-type stars with large residuals after fitting a zero point to transformed CCD observations. PEP data were ignored when determining periods for these stars.}
  \label{T2}
  \setlength{\tabcolsep}{2mm}
{\small
  \begin{tabular}{@{}rllllll}
  \hline
  HIP & Name & Amp   & Std Dev & Period & Spectral\\
      &      & (mag) & (mag)   & (days) & Type    \\
  \hline
   61404 & BO Mus    & 0.55 & 0.069 & 133  & M6II/III \\ % K156
   66666 & V744 Cen  & 1.11 & 0.122 & 167  & M8III \\    % K53
   98608 & $\nu$ Pav & 0.45 & 0.066 & 87.8 & M6III \\    % K29
  106044 & SX Pav    & 0.54 & 0.064 & 50.5 & M5III \\    % K15
  \hline
\end{tabular}
}
\end{table}

% Fig 3.4 [064/all.eps]

% Fig 3.7 [068/v1/k2lc.png]

\subsection{VZ Cam}

VZ Cam was observed by one of us (B.Sz.) using the 24$^{\prime \prime}$ telescope of the Konkoly Observatory between 1968 and 1977. The data were obtained with an RCA 1P21 photomultiplier through Johnson $UBV$ filters, using two comparison stars (comp=HD~57508, $V$=6.550, $B-V$=0.966 and $U-B$=0.63; check=HD~55075, $V$=6.325, $B-V$=$-$0.040 and $U-B$=$-$0.05). The typical photometric accuracy is $\pm$0.01--0.02 mag. The star was observed in a program for obtaining long-term photometric data for the peculiar Pop. II Cepheid RU~Cam \citep{b_sze}. We added the object to our sample because of its accurate Hipparcos parallax and the previously unknown two periods that can be determined from the decade-long light curve. This is the first analysis of VZ~Cam in the literature.

\section{Period Analysis}
\label{s_period_anal}

Characteristic periods were determined for each light curve using a weighted Discrete Fourier Transform (DFT). The weighted mean magnitude of the set was first subtracted to prevent a peak at zero frequency from dominating the spectrum. Observations were weighted by the inverse square of their uncertainty, except those combined with PEP data, which were analysed using multiple weighting schemes as described below. The DFT code was an implementation of the algorithm proposed by \citet[][Appendix D]{b_sca}, modified to include weights for each point.

Observations were predominantly spaced at intervals of 1 day, resulting in a Nyquist frequency of 0.5 cycle day$^{-1}$.  Power spectra were created for the frequency range 0--0.4 cycle day$^{-1}$. When searching for the most significant peak (greatest power), only frequencies $< 0.2$ cycle day$^{-1}$ were considered, corresponding  to a pulsation period $>$ 5 d. %, well below the acoustic cut-off frequency observed in the Magellanic Clouds \citep{b_sos07}.
The mean power in the frequency range 0.2--0.4 cycle day$^{-1}$, where no pulsation signal is expected, was used to estimate the noise level. The peaks at low frequency have lower S/N because of the rising noise floor, caused by $1/f$ noise. Only periods shorter than 300 d were kept, since we consider them to be the most reliable. Longer periods were less well determined due to gaps in the sampling, possible instrumental drifts, and difficulty matching light curves in different phases due to insufficient comparison stars in the earlier, smaller FOVs.

Light curves containing both CCD and PEP data showed higher noise in their corresponding DFTs compared to those prepared using CCD data only. Noise levels were at a minimum when all observations were weighted equally, but increased for weighting schemes using $\sigma^{-1}$ and $\sigma^{-2}$. This effect was caused by the $V$-filtered PEP observations having larger amplitudes than the unfiltered CCD data (Figure~\ref{fig011}), probably caused by the lack of instrumental colour information in the transformation. However, since the periods are well-correlated, we find that despite the increase in noise, the fitted periods remain unaffected. Since equal numbers of PEP and CCD observations tend to produce the greatest noise, the weighting scheme was selected dynamically based on the relative number of observations. An inverse variance was used when the number of CCD observations exceeded the PEP observations by a factor of two, otherwise equal weights were used.

Periods were extracted using iterative sinewave fitting \citep{b_fra}. That is, least-squares minimisation was used to fit a sinusoid with a frequency corresponding to the strongest peak in the DFT, while solving for the optimum amplitude and phase. The best-fitting sinusoid was subtracted from the data, and the process repeated until no strong peaks (S/N $\ge 7.5$) remained.

\subsection{The catalogue}
\label{s_results}

The survey results\bold{\footnote{Available electronically from CDS, Strasbourg.}} are presented in Table \ref{T4}, with stars ordered by ascending Hipparcos catalogue number (column 1). Columns 2 and 3 list the star name, and spectral type from the Bright Star Catalogue \citep{b_hof91} where available, otherwise from the General Catalogue of Variable Stars \citep[GCVS]{b_sam}, or the most recent entry from the compilation of spectral types by \citet{b_ski}. Additional columns list the Johnson $V$-band magnitude, revised Hipparcos parallax and uncertainty \citep{b_van07}, apparent $K$ magnitude, absolute $K$ magnitude and uncertainty, $K$-band magnitude source (described in Section \ref{s_k_mags}; D=DIRBE, G=Gezari, 2M=2MASS), number of \bold{CCD and PEP} observations, and detected periods.

\bold{GZ Eri and NSV 16260 are listed as G and K giants, respectively, although other sources list them as M giants \citep{b_hou88,b_hou82}. The shape of their light curves and the derived periods are consistent with the usual variability in early M-type stars.} Fourteen stars have no determined periods, although we retain them in Table \ref{T4} to document their $M_K$. Five of these stars (BW Oct, NSV 18694, V931 Her, V983 Her, and V2158 Cyg)
% V931 Her  +35 25  K058 x
% NSV 18694 -86     K106 x
% V983 Her  +29 10  K123 x
% V2158 Cyg +35 50  K146 x
% BW Oct    -84     K199 x
have no observations due to physical obscuration or low altitude. A further 6 stars ($\gamma$ Hyi, $\lambda$ Vel, V1919 Cyg, V2090 Cyg, V2114 Cyg, V345 Peg) have insufficient observations for reliable period determination.
% V2090 Cyg,    K087 x
% V345 Peg,     K094 x
% V2114 Cyg,    K164 x
% V1919 Cyg     K170 x
% lambda Vel    K253 x
% gamma Hyi     K261 x
The remaining 3 stars have either long periods, or periods with low S/N, both of which were excluded by our selection criteria, as follows. X Tra shows two periods of $\sim385$ and $\sim 455$ d. CI Boo shows a long period of $\sim 485$ d and a shorter period of $\sim 51$ d, although the latter is only detected at S/N $\approx$ 7.2. V440 Aur has a possible 39 d period which is only detected at S/N $\approx$ 5.8.

\subsection{Sample light curves}

%A selection of light curves and their corresponding power spectra are plotted in Figures \ref{fig019}--\ref{fig023}. The full set of plots \bold{are available on request from the authors, and photometric data are available electronically from CDS, Strasbourg.}

A selection of light curves and their corresponding power spectra \bold{are shown in Appendix A in the online version of the paper (see Supporting Information). Figures \ref{fig019} and \ref{fig019p} show an example light curve and power spectrum for Z Eri. The full set of plots are available on request from the authors, and photometric data are available electronically from CDS, Strasbourg.}

Relatively few stars show Mira-like light curves with regular, high-amplitude, mono-periodic behaviour, although our selection in the appendix is biased toward higher-amplitude variability for aesthetic reasons (see for example, DU Cha and AY Dor). Instead, the majority of stars exhibit short-period variability on time-scales of 10--100 d, small amplitudes ($A <$ 0.2 mag), and multiple periods, resulting in complicated light curves (see, for example, V512 Car).

%The light curve of Z Eri shows a significant change in amplitude around JD 2453900, together with a small change in period. This is also reflected in its DFT which shows a broad peak around 0.013 cycle d$^{-1}$, consisting of two closely spaced frequencies ($f_0$=0.0133 c/d, $f_1$=0.0125 c/d). The cross-production term, $f_0$+$f_1$, is also visible.

It is well known that the excited frequencies in SRVs are not stable over time, and may vary by a few percent \citep{b_mat97,b_kis99}. The amplitudes of the modes can vary on timescales of 1--5 yr \citep{b_kis00,b_per03}, and in some cases, phase variations have been observed on short time-scales, consistent with the properties of a stochastically excited and damped oscillator \citep{b_bed05}. Consequently, the complex light variations of SRVs cannot be fully modelled by a simple sum of sine waves. The occurrence of closely separated frequencies is quite common amongst our sample (see, for example, Z Eri, WX Hor, $\epsilon$ Mus, BL Cru), and is probably caused by modulations in the amplitude and/or phase of a single oscillation mode, and may indicate the finite lifetime of the oscillation. We did not try to combine periods by introducing arbitrary criteria, but instead, provide the raw data in Table \ref{T4}.

We found that 63\% of periods are shorter than 50 d and nearly 5\% of periods are shorter than 15 d. Of these, the shortest reliable period is 8.7 d (NSV 15560, $A=$9 mmag) and corresponds to the strongest peak in its power spectrum (see Fig %\ref{fig023}).
\bold{A6).}
The shortest period belongs to NSV 18346 ($P$=6.7 d, M3III, $J-K$=1.16, $M_K$=$-$4.7) but has an amplitude of only 5 mmag and may be spurious.

Although we did not retain periods longer than 300 d in Table \ref{T4}, we note that WW Pic exhibits a long period of $\sim$ 370 d and a short period of 36.1 d. The period ratio of $\sim$ 10 is consistent with galactic SRVs having long secondary periods (LSP). Although the precise nature of their variations are still unknown \citep{b_oli03,b_woo04}, LSP behaviour has been identified amongst SRVs in the lower metallicity environment of the LMC \citep{b_woo99}. Moreover, WW Pic has an $M_K$=$-$4.74, placing the 370-d period on sequence D of \citet{b_woo99}, which is composed of LMC stars having LSPs.

\begin{comment} %%%%%%%%%%%%%%%%%%%%%%%% one page %%%%%%%%%%%%%%%
\begin{table*}
  \caption{Survey results.}
  \label{T4}
{\small
  \begin{tabular}{@{}rlllrrrrrrlrl}
  \hline
 HIP & Name & Spectral & Var & $m_V$ & $\pi$ & $\sigma_{\pi}$ & $m_K$ & $M_K$ & $\sigma_{M_K}$ & Src & N & Periods\\
     &      & Type     &     & (mag) & (mas) & (mas)          &  (mag)& (mag) & (mag)          &     &   & (days) \\
  \hline
  \input{pl.inc}
  \hline
\end{tabular}
}
\end{table*}
\end{comment}

%%%%%%%%%%%%%%%%%%%%%%%%%%%%%%%%%%%%%%%% multi-page %%%%%%%%%%%%%%%%%%%%%%%%
\begin{scriptsize}

\begin{center}
\onecolumn
\begin{longtable}{rllrrrrlrrl}
\caption{Survey results.}\label{T4} \\
\hline
% HIP & Name & Spectral & Var & $m_V$ & $\pi$ & $\sigma_{\pi}$ & $m_K$ & $M_K$ & $\sigma_{M_K}$ & Src & N & Periods\\
%     &      & Type     &     & (mag) & (mas) & (mas)          &  (mag)& (mag) & (mag)          &     &   & (days) \\

%  HIP & Name & Spectral & Var & $m_V$ & $\pi$ & $m_K$ & $M_K$ & Src & N & Periods\\
%      &      & Type     &     & (mag) & (mas) & (mag) & (mag) &     &   & (days) \\

  HIP & Name & Spectral & $m_V$ & $\pi$ & $m_K$ & $M_K$ & Src & $N_C$ & $N_P$ & Periods\\
      &      & Type     & (mag) & (mas) & (mag) & (mag) &     &       &       & (days) \\
\hline
\endfirsthead
\hline
% HIP & Name & Sp & Var & $m_V$ & $\pi$ & $\sigma_{\pi}$ & $m_K$ & $M_K$ & $\sigma_{M_K}$ & Src & N & Periods\\
%     &      &    &     & (mag) & (mas) & (mas)          &  (mag)& (mag) & (mag)          &     &   & (days)  \\

%  HIP & Name & Spectral & Var & $m_V$ & $\pi$ & $m_K$ & $M_K$ & Src & N & Periods\\
%      &      & Type     &     & (mag) & (mas) & (mag) & (mag) &     &   & (days) \\

  HIP & Name & Spectral & $m_V$ & $\pi$ & $m_K$ & $M_K$ & Src & $N_C$ & $N_P$ & Periods\\
      &      & Type     & (mag) & (mas) & (mag) & (mag) &     &       &       & (days) \\

\hline
\endhead
%This is the footer for all pages except the last page of the table...
 \multicolumn{11}{l}{{Continued on next page\ldots}} \\
% \hline \\
\endfoot
%This is the footer for the last page of the table...
  \\ \hline
\endlastfoot

%\multicolumn{13}{l}{} \\
     154&YY Psc          &M3III     & 4.37& 7.55 (0.59)&$-$0.47&$-$6.09 (0.17)& D&267&  0&23.1 32.0 53.6 167.8\\
  1158&AD Cet          &M3+III    & 5.13& 5.60 (0.30)&$+$0.50&$-$5.77 (0.12)& D&272&  0&18.6 29.2 212.3\\
  1170&AE Cet          &M3III     & 4.44& 7.29 (0.28)&$+$0.20&$-$5.50 (0.08)& D&324&  0&19.2 19.6 27.1 41.7\\
  2086&HD 2268         &M1III     & 6.24& 2.07 (0.66)&$+$1.74&$-$6.69 (0.69)& G&260&  0&16.8 22.7 39.1 202.0\\
  2210&$\eta$ Scl      &M4III     & 4.86& 7.22 (0.51)&$+$0.09&$-$5.62 (0.15)& D&334&  0&22.7 23.5 24.6 47.3 128.7 158.7\\
  2215&AG Cet          &M3        & 7.35& 4.22 (0.56)&$-$0.04&$-$6.92 (0.29)& G&273&  0&57.6 126.3\\
  2219&TV Psc          &M3III     & 5.01& 6.17 (0.59)&$-$0.17&$-$6.23 (0.21)& D&207&  0&55.1 216.5 266.7\\
  3632&NSV 00293       &M4III     & 5.36& 4.20 (0.29)&$+$0.25&$-$6.65 (0.15)& D&205&  0&24.3 25.7 37.3 115.2\\
  3894&BU Phe          &M4III     & 7.19& 3.70 (0.45)&$+$2.00&$-$5.17 (0.27)& D&420&  0&22.7 32.9 92.9 218.3\\
  4200&BQ Tuc          &M4III     & 5.73& 4.58 (0.28)&$+$0.20&$-$6.50 (0.13)& D&454&138&31.6 45.0 60.9\\
  4317&HR 259          &M4III     & 6.19& 5.75 (0.42)&$+$1.04&$-$5.18 (0.16)& D&137&  0&18.8\\
  4879&CC Tuc          &M2III     & 6.25& 3.76 (0.32)&$+$1.57&$-$5.56 (0.19)& D&425&161&16.2 20.3 20.6\\
  6487&NSV 00490       &M3III     & 7.05& 3.40 (0.69)&$+$2.52&$-$4.83 (0.45)& D&252&  0&16.4 18.3 34.6\\
  6952&AW Phe          &M2III     & 6.25& 3.22 (0.42)&$+$1.35&$-$6.11 (0.29)& D&387&  0&30.5 43.0 66.0 71.5 159.0 219.3\\
  7505&CY Psc          &M4        & 6.39& 3.54 (0.49)&$+$1.53&$-$5.73 (0.30)& D&210&  0&33.5\\
  7506&NSV 15347       &M3III     & 6.18& 4.37 (0.32)&$+$1.41&$-$5.39 (0.16)& D&429&  0&13.9 17.9 23.6 37.6 91.7\\
  8240&HD 10934        &M3III     & 5.49& 5.14 (0.25)&$+$1.04&$-$5.41 (0.11)& D&418&  0&11.8 13.8 20.0 20.2 158.7\\
  8837&$\psi$ Phe      &M4III     & 4.39& 9.54 (0.19)&$-$0.65&$-$5.75 (0.04)& D&376&119&43.7 48.1\\
  9171&AA Tri          &M5.5      & 6.94& 5.11 (0.46)&$+$0.17&$-$6.29 (0.20)& D& 49&  0&23.5\\
  9372&AR Cet          &M3III     & 5.43& 6.85 (0.31)&$-$0.39&$-$6.21 (0.10)& D&248&  0&22.0 22.9 28.7 37.3 100.7\\
 10155&NSV 00738       &M3III     & 5.68& 5.84 (0.49)&$+$0.95&$-$5.25 (0.18)& D&123&  0&18.1 21.9\\
 11293&TZ Hor          &M5III     & 6.39& 4.33 (0.41)&$+$0.78&$-$6.05 (0.21)& D&405&131&23.3 52.9 69.8 116.7\\
 11455&CL Hyi          &M6/7      & 7.17& 6.20 (0.45)&$+$0.03&$-$6.02 (0.16)& D&392&  0&66.4 71.9 75.8 94.1 100.1\\
 11648&TV Hor          &M4-5III   & 6.76& 3.42 (0.33)&$+$1.11&$-$6.24 (0.21)& D&424&  0&32.9 33.6 34.8 47.1 69.9 248.1\\
 12016&FH Eri          &M4III     & 7.03& 3.22 (0.51)&$+$1.65&$-$5.82 (0.35)& D&375&  0&20.0 20.8 21.7 22.1 28.5 30.2 156.7\\
 12433&NSV 15560       &M3III     & 7.85& 2.56 (0.59)&$+$3.06&$-$4.91 (0.52)& D&360&  0&8.7 112.2\\
 13064&Z Eri           &M4III     & 6.63& 3.90 (0.49)&$+$0.34&$-$6.71 (0.27)& D&253&  0&75.4 80.1 123.5\\
 13613&NSV 15611       &M3III     & 7.38& 4.25 (0.68)&$+$2.80&$-$4.06 (0.37)& D&323&  0&12.2 23.5 28.5 253.8\\
 13654&RZ Ari          &M6-III    & 5.76& 9.28 (0.30)&$-$1.03&$-$6.23 (0.07)& D&141&  0&49.9 54.8\\
 13756&EH Cet          &M4III     & 6.15& 4.42 (0.53)&$+$1.11&$-$5.69 (0.26)& D&196&  0&37.2 62.5 91.6\\
 14456&HD 19349        &M3III     & 5.23& 7.15 (0.41)&$+$0.52&$-$5.22 (0.13)& D&232&  0&12.9 31.8 115.7 191.9\\
 14930&TW Hor          &C7,2(N0)  & 5.71& 3.11 (0.37)&$+$0.09&$-$7.45 (0.26)& D&  0& 45&270.3\\
 15474&$\tau^4$ Eri    &M3.5III   & 3.70&10.71 (0.54)&$-$1.19&$-$6.04 (0.11)& D&284&  0&23.8\\
 16484&NSV 15716       &M2III     & 7.92& 2.45 (0.50)&$+$3.62&$-$4.45 (0.52)&2M&398&  0&18.5 25.5\\
 17678&$\gamma$ Hyi    &M2III     & 3.26&15.24 (0.11)&$-$0.93&$-$5.02 (0.03)& D&  0& 16&\\
 17889&WX Hor          &M5/6III   & 7.45& 3.13 (0.54)&$+$0.67&$-$6.86 (0.38)& D&382&  0&66.8 92.6 99.8 156.3 260.4\\
 17993&GL Eri          &M5III     & 7.05& 3.24 (0.44)&$+$1.38&$-$6.08 (0.30)& D&322&  0&29.6 39.6 41.8 285.7\\
 18048&GM Eri          &M4III     & 7.57& 3.37 (0.58)&$+$1.77&$-$5.60 (0.38)& D&336&  0&60.4 123.0 210.1\\
 18605&NSV 15865       &M3III     & 7.64& 2.81 (0.43)&$+$3.15&$-$4.62 (0.37)& D&378&  0&94.2 161.8\\
 18691&XY Dor          &M1III     & 6.51& 2.61 (0.38)&$+$1.99&$-$5.92 (0.33)& D&381&  0&13.3 17.1 22.9 28.9 190.8 271.7\\
 18744&$\gamma$ Ret    &M4III     & 4.48& 6.95 (0.11)&$-$0.45&$-$6.24 (0.04)& G&383&157&30.0 42.8 277.0\\
 18931&DP Eri          &M5III     & 7.24& 3.46 (0.66)&$+$1.64&$-$5.67 (0.42)& D&310&  0&15.5 26.8 90.5 183.2\\
 20075&GZ Eri          &G8III     & 6.00& 2.16 (0.38)&$+$0.72&$-$7.62 (0.38)& D&320&  0&49.1 52.0 57.4 224.7\\
 20856&RV Cae          &M1III     & 6.41& 2.74 (0.35)&$+$2.15&$-$5.67 (0.28)& D&334&  0&25.7 34.5 37.5 117.2\\
 20958&V1145 Tau       &M3.5III   & 6.93& 3.59 (0.60)&$+$1.58&$-$5.70 (0.37)& D&189&  0&27.2 41.3 84.7\\
 21296&DV Eri          &M3III     & 5.20& 3.28 (0.26)&$+$0.66&$-$6.79 (0.17)& D&234&  0&28.9 44.2 62.2\\
 21479&R Dor           &M8III     & 5.59&18.31 (0.99)&$-$4.02&$-$7.71 (0.14)& D&  0& 72&168.9\\
 21763&DM Eri          &M4III     & 4.32& 8.85 (0.60)&$-$0.34&$-$5.61 (0.15)& D&304&  0&18.8 45.5\\
 22667&$o^1$ Ori       &S3.5/1-   & 4.71& 5.01 (0.71)&$-$0.51&$-$7.02 (0.31)& D&171&  0&30.8 70.7\\
 22737&NSV 1777        &M0-1III   & 6.43& 2.73 (0.37)&$+$2.34&$-$5.48 (0.41)&2M&335&  0&19.8 20.4 27.4 45.5 83.8 95.2\\
 23653&NSV 16242       &M2III     & 6.26& 3.95 (0.35)&$+$2.03&$-$4.99 (0.21)& D&367&  0&13.1 18.4 92.9 109.4\\
 23840&WZ Dor          &M3III     & 5.19& 5.77 (0.18)&$+$0.33&$-$5.87 (0.07)& D&388&135&26.0 44.5\\
 23874&NSV 01842       &M1III     & 6.33& 3.89 (0.45)&$+$1.74&$-$5.35 (0.26)& D&255&  0&13.4 16.7 17.3 42.8\\
 24169&RX Lep          &M6III     & 5.60& 6.71 (0.44)&$-$1.38&$-$7.26 (0.14)& D&248&  0&90.1 101.7\\
 24189&NSV 16260       &K5III     & 6.57& 2.44 (0.39)&$+$2.42&$-$5.65 (0.36)& D&298&  0&15.9 22.3 30.5 66.4 90.7\\
 24441&V1365 Ori       &M3        & 6.89& 4.10 (0.53)&$+$0.71&$-$6.24 (0.28)& D&220&  0&42.3 62.7\\
 24943&WW Pic          &M3/4III   & 6.60& 2.94 (0.39)&$+$1.92&$-$5.74 (0.29)& D&334&  0&36.1 72.8 285.7\\
 25194&SW Col          &M1III     & 5.73& 5.02 (0.37)&$+$1.14&$-$5.37 (0.16)& D&316&  0&78.7 156.5\\
 26005&NSV 16342       &M2/3III   & 7.07& 2.34 (0.53)&$+$2.59&$-$5.57 (0.50)& D&303&  0&18.1 20.2 27.0 39.2 91.0\\
 26169&WX Men          &M3III     & 5.79& 3.00 (0.31)&$+$0.60&$-$7.04 (0.23)& D&283&170&38.5 39.7 40.1 43.5 54.4 115.2\\
 27229&WY Lep          &M4/M5III  & 6.54& 5.28 (0.47)&$+$0.59&$-$5.80 (0.19)& D&288&  0&32.2 47.3 162.9 183.8\\
 27344&NSV 16680       &M5III     & 6.78& 3.31 (0.35)&$+$1.87&$-$5.54 (0.24)& D&324&  0&14.2 14.7 21.2 50.2 196.5\\
 27661&V440 Aur        &M3III     & 6.28& 5.44 (0.72)&$+$1.17&$-$5.17 (0.29)& D& 46&  0&\\
 28596&SW Pic          &M4III     & 6.41& 3.40 (0.32)&$+$1.13&$-$6.22 (0.21)& D&396&123&25.1 35.5\\
 28874&S Lep           &M6III     & 6.74& 4.92 (0.63)&$-$0.55&$-$7.10 (0.28)& D&280&  0&97.3\\
 28984&YY Lep          &M4III     & 5.74& 2.60 (0.39)&$+$0.46&$-$7.47 (0.33)& D&282&  0&38.8 41.6 56.9 80.6\\
 29048&NSV 16782       &M2III     & 5.28& 3.89 (0.33)&$+$1.03&$-$6.03 (0.19)& D&300&  0&19.9 29.2 39.6 88.7 165.6\\
 29263&AF Col          &M2II-III  & 5.54& 3.83 (0.25)&$+$0.66&$-$6.44 (0.15)& D&319&  0&42.0 43.9 48.6 112.5 250.6\\
 29353&$\eta^2$ Dor    &M2.5III   & 5.01& 5.09 (0.23)&$+$0.45&$-$6.03 (0.10)& D&337& 46&21.4 21.8 23.3 150.2\\
 29509&V1389 Ori       &M2III     & 6.45& 5.48 (0.51)&$+$2.12&$-$4.20 (0.22)& D&218&  0&18.0 27.8 35.2\\
 30237&AC Pic          &M3III     & 6.79& 2.61 (0.40)&$+$2.01&$-$5.91 (0.35)& D&374&  0&25.8 26.2 35.0\\
 30343&$\mu$ Gem       &M3III     & 2.87&14.08 (0.71)&$-$1.88&$-$6.15 (0.11)& D&154&  0&25.2 40.5\\
 30352&NSV 16859       &M1III     & 6.30& 2.96 (0.29)&$+$1.90&$-$5.75 (0.22)& D&358&  0&15.3 19.8\\
 31057&AY Dor          &M6II/III  & 7.22& 3.03 (0.47)&$+$0.60&$-$6.99 (0.34)& G&369&  0&78.3 86.4\\
 31086&KM CMa          &M4III     & 6.31& 3.18 (0.41)&$+$1.57&$-$5.94 (0.28)& D&303&  0&57.6 115.6 234.7\\
 31099&SX Col          &M1III     & 6.26& 4.53 (0.31)&$+$1.50&$-$5.23 (0.16)& D&314&  0&17.5 21.2 22.0 22.5 30.3 31.5 89.6\\
 31363&V729 Mon        &M2III     & 6.68& 3.75 (0.67)&$+$1.52&$-$5.63 (0.49)&2M&227&  0&23.3 33.5 47.5\\
 32374&NSV 17193       &M2III     & 7.18& 2.61 (0.47)&$+$2.59&$-$5.32 (0.41)& D&365&  0&14.6 15.7 35.7 56.0\\
 33040&KX CMa          &M4III     & 6.33& 3.58 (0.46)&$+$0.42&$-$6.82 (0.28)& D&271&  0&37.1 50.6 62.7\\
 35120&NSV 03486       &M4-III    & 5.78& 6.01 (0.37)&$+$0.74&$-$5.37 (0.14)& G&215&  0&11.9 30.2 64.7\\
 35626&MZ CMa          &M4III     & 5.87& 5.10 (0.37)&$+$1.21&$-$5.26 (0.16)& G&296&  0&15.8 17.6 18.7 23.9 25.6 34.9 49.6\\
 35746&GH CMa          &M6III     & 6.96& 3.31 (0.67)&$+$0.84&$-$6.58 (0.44)& D&295&  0&44.4 55.7 62.1 89.7\\
 36547&VZ Cam          &M4III     & 4.92& 6.53 (0.25)&$-$0.13&$-$6.05 (0.19)&2M&  0&695&27.1 28.1 38.5 39.0 54.4 205.3 249.4\\
 36982&NSV 17548       &M0III     & 6.72& 2.69 (0.36)&$+$2.07&$-$5.78 (0.31)& D&135&  0&13.2\\
 37321&NSV 17542       &M1III     & 7.38& 2.96 (0.63)&$+$2.59&$-$5.08 (0.50)&2M&256&  0&21.8 49.1 184.8\\
 37508&NSV 17555       &M2III     & 5.79& 4.19 (0.33)&$+$1.34&$-$5.56 (0.18)& G&218&  0&13.2 18.4 37.7\\
 37521&NZ Gem          &M3II-III  & 5.55& 2.52 (0.60)&$+$0.71&$-$7.29 (0.52)& D&200&  0&30.1 31.5 43.8 63.7 166.4\\
 38406&BC CMi          &M4III     & 6.30& 6.45 (0.47)&$+$0.79&$-$5.17 (0.16)& D&230&  0&27.7 143.3 208.3\\
 38834&V341 Car        &M0III     & 7.06& 3.81 (0.40)&$+$1.29&$-$5.83 (0.23)& D&396&  0&30.5 274.7\\
 39070&V460 Car        &M1.5IIa   & 5.19& 3.32 (0.20)&$+$0.68&$-$6.74 (0.35)&2M&413&  0&32.9 34.3\\
 39299&NSV 03872       &M1III     & 6.31& 4.06 (0.51)&$+$1.38&$-$5.62 (0.36)&2M&331&  0&21.2 25.4\\
 39659&BL Cnc          &M3III     & 5.96& 4.38 (0.53)&$+$1.30&$-$5.50 (0.27)& D&176&  0&22.6 37.8 203.7\\
 40144&V427 Pup        &M5        & 6.64& 3.77 (0.34)&$+$1.00&$-$6.17 (0.34)&2M&307&  0&38.7 52.5 57.4 243.3\\
 40282&HR 3247         &M1III     & 5.52& 4.52 (0.22)&$+$2.00&$-$4.75 (0.11)& G&366&  0&13.2 16.1 24.3 114.9\\
 41080&NSV 17870       &M1III     & 5.92& 2.53 (0.59)&$+$1.41&$-$6.58 (0.51)& D&247&  0&17.7 22.8 25.5 33.4 94.5 220.3\\
 42176&NSV 17929       &M0        & 7.36& 4.66 (0.77)&$+$2.86&$-$3.81 (0.40)& D&240&  0&37.9\\
 42502&AK Hya          &M4III     & 6.59& 6.40 (0.41)&$-$0.61&$-$6.59 (0.14)& D&294&  0&78.6 88.7 133.7\\
 43215&AK Pyx          &M5III     & 6.42& 4.68 (0.41)&$+$0.57&$-$6.09 (0.19)& D&340&  0&55.5 57.9 86.7 162.9 232.6\\
 44126&FZ Cnc          &M4III     & 6.30& 5.32 (0.44)&$+$1.23&$-$5.14 (0.18)& D&210&  0&17.6\\
 44281&AL Pyx          &M1III     & 7.19& 2.70 (0.42)&$+$2.39&$-$5.48 (0.34)& G&308&  0&36.1 44.5 55.3 60.0 192.3 284.1\\
 44538&NSV 18130       &M4/5III   & 7.78& 3.55 (0.54)&$+$2.21&$-$5.06 (0.35)& D&349&  0&9.5 15.2 15.6 140.1\\
 44816&$\lambda$ Vel   &K4Ib-IIa  & 2.23& 5.99 (0.11)&$-$1.57&$-$7.68 (0.04)& D&  0& 26&\\
 45902&$\theta$ Pyx    &M1III     & 4.71& 6.49 (0.45)&$+$0.54&$-$5.41 (0.15)& D&338&  0&13.0 98.3\\
 45924&NSV 18181       &M1III     & 5.56& 5.68 (0.28)&$+$1.23&$-$5.01 (0.33)&2M&319&  0&15.1 21.7 32.0 123.0 143.7\\
 46027&NSV 18186       &M2III     & 6.83& 2.54 (0.54)&$+$1.73&$-$6.27 (0.47)& D&286&  0&15.7 15.9 20.6 27.5 130.9\\
 46521&AL Ant          &M2/M3III  & 7.48& 3.06 (0.65)&$+$2.39&$-$5.20 (0.47)& D&390&  0&27.5 38.2 40.8 131.4 174.2\\
 46620&V482 Car        &M2III     & 5.88& 3.37 (0.45)&$+$1.02&$-$6.36 (0.38)&2M&416&  0&25.5 26.0 34.2 166.1 216.9\\
 47608&NSV 18257       &M1/2III   & 6.97& 2.92 (0.48)&$+$2.57&$-$5.14 (0.44)&2M&396&  0&8.7 11.2 14.5 19.2 67.0 97.7\\
 49610&IO Hya          &M4III     & 6.99& 3.62 (0.72)&$+$1.40&$-$5.82 (0.43)& D&347&  0&26.3 31.9 34.5 43.9 93.7 170.1\\
 49755&PS Hya          &M3III     & 7.09& 1.59 (0.78)&$+$2.40&$-$6.62 (1.07)& D&308&  0&17.4 29.6 99.2\\
 49902&NSV 18346       &M3III     & 6.74& 4.56 (0.48)&$+$1.98&$-$4.74 (0.24)& D&332&  0&6.7 8.9 141.6\\
 49926&V368 Car        &M3III     & 6.07& 4.07 (0.30)&$+$1.05&$-$5.92 (0.30)&2M&454&  0&22.6 29.4 30.0 41.6\\
 50332&GY Vel          &M4-5III   & 6.20& 4.56 (0.33)&$+$0.42&$-$6.30 (0.16)& G&415&  0&20.0 25.3 25.9 26.5 35.7 36.7 116.6\\
 50626&V501 Car        &M4III     & 6.96& 3.33 (0.36)&$+$1.70&$-$5.70 (0.36)&2M&314&  0&22.7 24.2 32.4 45.8 48.2\\
 50827&UV Sex          &M3III     & 7.05& 3.10 (0.81)&$+$2.08&$-$5.48 (0.57)& D&305&  0&34.6 36.6 48.7 83.8\\
 51049&V347 Vel        &M1III     & 7.00& 2.79 (0.45)&$+$2.60&$-$5.21 (0.43)&2M&406&  0&19.5 29.1 32.2 42.6 46.3 136.6 179.5\\
 51141&V505 Car        &M3III     & 6.61& 3.15 (0.29)&$+$1.79&$-$5.74 (0.36)&2M&259& 61&26.5\\
 51327&NSV 18423       &M1III     & 6.42& 4.90 (0.29)&$+$1.95&$-$4.62 (0.19)& D&435&  0&9.1 11.7 23.6 182.1\\
 51810&V512 Car        &M3III     & 6.87& 2.76 (0.41)&$+$1.92&$-$5.90 (0.34)& D&423&  0&25.2 34.6 35.7 50.2\\
 51905&$\phi^2$ Hya    &M1III     & 6.01& 4.31 (0.33)&$+$1.87&$-$4.97 (0.18)& D&334&  0&11.0 110.3 153.6\\
 53449&VY Leo          &M5.5III   & 5.91& 8.39 (0.37)&$-$0.79&$-$6.18 (0.10)& D&236&  0&35.8 54.9 75.0 84.5\\
 54474&NSV 18694       &M2III     & 6.87& 2.76 (0.36)&$+$2.18&$-$5.65 (0.30)& D&  0&  0&\\
 54593&V897 Cen        &M3III     & 7.47& 3.12 (0.61)&$+$2.23&$-$5.32 (0.49)&2M&395&  0&16.2 21.2 21.4 30.3 39.2 82.8\\
 54708&V532 Car        &M4III     & 6.24& 4.12 (0.42)&$+$0.65&$-$6.29 (0.23)& D&463&  0&33.7 35.3 49.6 91.0\\
 54725&NSV 05129       &M1III     & 6.35& 3.55 (0.47)&$+$1.78&$-$5.49 (0.29)& D&389&  0&15.0 153.4\\
 54974&UU Crt          &M4III     & 6.86& 1.84 (0.49)&$+$1.58&$-$7.12 (0.58)& D&312&  0&35.5 54.0 55.7 78.0 189.4 273.2\\
 54999&SY Crt          &M4III     & 6.51& 4.09 (0.73)&$+$1.04&$-$5.92 (0.39)& D&318&  0&28.6 32.4 38.7 42.5 45.6 176.4\\
 55238&V537 Car        &M6III     & 6.61& 2.97 (0.50)&$-$0.05&$-$7.71 (0.52)&2M&461&  0&39.9 51.2 52.8 71.9\\
 56293&NSV 18788       &M1III     & 6.15& 4.54 (0.36)&$+$1.57&$-$5.15 (0.18)& D&390&  0&16.7 18.3 24.6 28.5 34.9 133.0\\
 56698&V912 Cen        &M3/M4III  & 6.85& 2.19 (0.62)&$+$1.65&$-$6.68 (0.62)& D&338&  0&27.9 37.3 38.4 40.1 137.4\\
 56702&V913 Cen        &M5/7      & 6.56& 6.08 (0.52)&$-$0.06&$-$6.15 (0.19)& G&443&  0&34.0 70.6 247.5\\
 56779&$\omega$ Vir    &M4III     & 5.24& 6.56 (0.36)&$-$0.24&$-$6.16 (0.12)& D&279&  0&23.9 25.0 29.5 31.0 59.0\\
 56970&V914 Cen        &M1III     & 5.95& 2.86 (0.52)&$+$1.67&$-$6.07 (0.48)&2M&361&  0&21.8 31.1 43.6 51.6 252.5\\
 57380&$\nu$ Vir       &M1III     & 4.04&11.10 (0.18)&$+$0.09&$-$4.69 (0.04)& D&264&  0&11.1 12.3 16.8 23.7\\
 57411&WW Crt          &M4III     & 7.54& 3.86 (0.59)&$+$1.97&$-$5.11 (0.33)& G&399&  0&16.0 17.7 23.7 26.4 50.9 83.5 123.3 133.9\\
 57505&DU Cha          &M6III     & 7.67& 4.42 (0.39)&$+$0.34&$-$6.45 (0.19)& D&323&  0&58.0 60.1\\
 57607&V919 Cen        &M7III     & 7.02& 4.14 (0.73)&$-$1.18&$-$8.11 (0.38)& D&405&  0&97.9 116.4 193.4 225.7\\
 57613&II Hya          &M4+III    & 5.10& 5.98 (0.27)&$-$0.30&$-$6.43 (0.10)& D&381&  0&29.5 30.4 39.5 41.4 89.1\\
 57800&RU Crt          &M3        & 8.14& 7.57 (0.98)&$+$0.07&$-$5.54 (0.28)& D&328&  0&60.6 66.4 101.1\\
 59588&V335 Hya        &M4III     & 6.34& 2.79 (0.46)&$-$0.17&$-$7.96 (0.36)& D&429&  0&107.6 141.2\\
 59929&$\epsilon$ Mus  &M5III     & 4.06&10.82 (0.17)&$-$1.43&$-$6.26 (0.03)& G&408&251&32.1 32.7 42.5 43.7 44.9 46.0 63.4 196.5\\
 60333&KO Vir          &M6        & 7.64& 4.53 (0.65)&$+$1.71&$-$5.02 (0.31)& D&314&  0&18.4 22.3 24.6 32.1 56.1 116.7 157.0\\
 60421&TT Crv          &M3III     & 6.48& 3.53 (0.39)&$+$1.64&$-$5.64 (0.24)& D&332&  0&15.4 15.6 19.5 19.9 26.8 37.7 182.8\\
 60781&BL Cru          &M4-5III   & 5.38& 7.36 (0.32)&$-$0.19&$-$5.87 (0.10)& D&482&179&30.7 42.3 43.6\\
 60979&V928 Cen        &M2II-III  & 6.00& 4.42 (0.37)&$+$1.16&$-$5.63 (0.19)& D&434&  0&19.0 19.2 22.3 24.6 29.6 46.4 102.2\\
 61084&$\gamma$ Cru    &M3.5III   & 1.59&36.83 (0.18)&$-$3.10&$-$5.27 (0.06)& G&209&232&12.1 15.1 16.5 54.8 82.7 104.9\\
 61404&BO Mus          &M6II-III  & 6.11& 2.97 (0.32)&$-$0.83&$-$8.48 (0.23)& D&488&156&134.8 229.9\\
 61658&FW Vir          &M3+III    & 5.68& 6.62 (0.31)&$+$0.94&$-$4.96 (0.10)& D&301&  0&9.3 28.6 89.7\\
 61908&V341 Hya        &M3III     & 6.64& 4.32 (0.43)&$+$1.72&$-$5.12 (0.23)& D&415&  0&8.6 18.3 18.7 36.2 216.5 281.7\\
 62247&LM Mus          &M5III     & 7.11& 4.87 (0.47)&$+$1.40&$-$5.18 (0.32)&2M&401&  0&20.8 28.9 30.0 30.5 42.4\\
 62985&$\psi$ Vir      &M3-III    & 4.77& 5.99 (0.23)&$+$0.14&$-$5.99 (0.09)& D&348&  0&22.4 23.5 24.5 30.1 31.3 49.4 162.6\\
 63090&$\delta$ Vir    &M3+III    & 3.39&16.44 (0.22)&$-$1.24&$-$5.16 (0.03)& D&229&  0&13.0 17.2 25.6 110.1 125.8\\
 63752&KX Vir          &M4.5III   & 7.41& 3.19 (0.71)&$+$1.76&$-$5.73 (0.49)& D&332&  0&15.6 17.8 23.4 25.6 69.4 77.6 250.0\\
 63950&FS Com          &M5III     & 5.53& 4.43 (0.41)&$-$0.20&$-$6.97 (0.20)& D&253&  0&33.7 44.6 46.8 159.5\\
 64117&NSV 06103       &M1III     & 6.34& 3.55 (0.44)&$+$1.81&$-$5.46 (0.41)&2M&313&  0&14.4 16.8 25.4 183.8 250.6\\
 64645&LO Vir          &M6        & 7.30& 3.75 (0.53)&$+$1.64&$-$5.50 (0.31)& D&332&  0&20.0 22.0 22.3 26.8 91.7\\
 64852&$\sigma$ Vir    &M1III     & 4.78& 4.83 (0.19)&$+$0.49&$-$6.10 (0.09)& D&328&  0&23.4 24.3 27.9 34.1 35.8 64.0 165.3\\
 65311&NSV 06212       &M1III     & 6.17& 3.99 (0.34)&$+$1.70&$-$5.31 (0.19)& D&440&  0&17.9 21.7 22.4 22.7 31.7 34.0 52.2 99.7\\
 66666&V744 Cen        &M5III     & 5.74& 6.35 (0.33)&$-$0.72&$-$6.72 (0.12)& D&461& 75&95.8 102.2 166.9\\
 66783&LY Mus          &M4III     & 6.91& 3.48 (0.41)&$+$0.95&$-$6.37 (0.26)& D&405&  0&53.1 55.3 84.7 106.6 128.7 261.8\\
 67288&NSV 06442       &M2III     & 5.41& 4.82 (0.39)&$+$1.01&$-$5.60 (0.18)& D&376&  0&18.3 18.5 18.8 25.4 25.9 50.2\\
 67457&V806 Cen        &M4.5III   & 4.19&17.82 (0.21)&$-$1.67&$-$5.41 (0.03)& D&354&107&26.5\\
 68815&$\theta$ Aps    &M7III     & 5.69& 8.84 (0.49)&$-$2.02&$-$7.29 (0.12)& D&  0&259&100.9 111.0\\
 68937&ER Vir          &M4III     & 6.57& 4.16 (0.50)&$+$1.45&$-$5.48 (0.26)& D&372&  0&16.1 18.8 25.7 29.6 80.2\\
 69269&ET Vir          &M2III     & 4.93& 7.08 (0.32)&$+$0.58&$-$5.18 (0.18)& D&367&  0&22.6 23.8 36.4 37.6 39.8 48.8 259.1\\
 69614&FS Vir          &M4III     & 6.41& 4.04 (0.52)&$+$1.54&$-$5.44 (0.28)& D&351&  0&13.5 16.1 17.0 22.6 144.7 170.1\\
 69850&NQ Aps          &M3III     & 7.28& 2.82 (0.62)&$+$1.93&$-$5.84 (0.49)& D&241&  0&21.2 28.2\\
 70236&CI Boo          &M3III     & 6.50& 4.23 (0.47)&$+$0.58&$-$6.29 (0.24)& D&141&  0&\\
 72122&HD 129902       &M1III     & 6.06& 3.46 (0.51)&$+$1.78&$-$5.54 (0.32)& D&283&  0&11.3 19.8 32.3 211.0\\
 72432&V768 Cen        &M3III     & 5.89& 6.34 (0.34)&$-$0.44&$-$6.44 (0.12)& D&460&  0&41.4 57.4 60.9 145.1 191.2\\
 72655&NSV 20194       &M0III     & 7.26& 3.83 (0.67)&$+$2.37&$-$4.73 (0.38)& G&515&  0&15.8 19.2 19.4 28.3 47.7 192.3\\
 73589&EV Boo          &M4III     & 6.60& 3.08 (0.41)&$+$0.70&$-$6.86 (0.29)& G&137&  0&31.2 35.3 59.3\\
 73764&GM Lup          &M6III     & 6.29& 5.40 (0.56)&$+$0.00&$-$6.35 (0.23)& G&387&  0&29.9 43.6 45.5 61.3 80.9 243.3\\
 74582&X TrA           &C5,5(Nb)  & 5.75& 2.78 (0.44)&$-$0.62&$-$8.39 (0.34)& D&  0&160&\\
 74982&FZ Lib          &M4III     & 6.95& 2.84 (1.03)&$+$1.00&$-$6.76 (0.79)& D&357&  0&22.2 25.9 26.4 62.3\\
 74999&OO Aps          &M2III     & 6.59& 2.82 (0.45)&$+$1.81&$-$5.93 (0.36)& D& 67&103&24.9\\
 76423&$\tau^4$ Ser    &M5IIb-III & 6.51& 4.86 (0.46)&$-$1.02&$-$7.60 (0.21)& D&329&  0&86.7 173.0 244.5\\
 76573&LY Ser          &M4III     & 6.82& 4.75 (0.47)&$-$0.46&$-$7.09 (0.22)& D&243&  0&66.7 76.5 220.8\\
 79072&FS Ser          &M3.5III   & 5.69& 4.04 (0.43)&$+$0.56&$-$6.41 (0.23)& G&354&  0&20.6 25.2 25.6 26.4 48.4\\
 79086&FQ Ser          &M4III     & 6.47& 4.92 (0.73)&$+$0.54&$-$6.01 (0.33)& G&354&  0&31.3 32.1 36.2 41.4 44.0 71.5 144.1\\
 79349&LQ Her          &M4.5III   & 5.74& 4.78 (0.31)&$+$0.15&$-$6.46 (0.14)& D&274&  0&24.5 25.1 28.0 29.9 33.6\\
 79754&V368 Nor        &M2III     & 5.45& 4.21 (0.43)&$+$0.62&$-$6.27 (0.22)& G&396&  0&18.9 21.4 22.1 24.2 33.6\\
 80047&$\delta^1$ Aps  &M5III     & 4.68& 4.28 (0.16)&$-$0.73&$-$7.59 (0.08)& G&307&  0&68.0 94.9 101.7\\
 80620&V2105 Oph       &M3-III    & 5.24& 5.65 (0.39)&$+$0.48&$-$5.80 (0.15)& D&371&  0&20.5 21.2 21.4 41.6 206.6\\
 81376&V840 Ara        &M2/M3III  & 6.84& 4.67 (0.57)&$+$1.84&$-$4.83 (0.38)&2M&384&  0&16.3 17.9 24.3 42.6 51.1 75.5 164.7\\
 81426&V904 Her        &M2III     & 6.96& 2.87 (0.52)&$+$2.19&$-$5.54 (0.40)& D&224&  0&22.8 65.5 95.4 257.7\\
 82028&HR 6227         &M3III     & 5.60& 7.19 (0.42)&$+$0.82&$-$4.91 (0.13)& D&345&  0&13.2 14.7 17.1 21.9 38.7\\
 83258&NSV 20880       &M4III     & 7.39& 3.05 (0.64)&$+$1.74&$-$5.86 (0.46)& G&345&  0&13.3 14.6 15.8 18.9 181.2\\
 83462&V931 Her        &M4III     & 6.19& 4.35 (0.33)&$+$0.70&$-$6.11 (0.17)& G&  0&  0&\\
 84744&NSV 21509       &M1III     & 6.80& 3.42 (0.48)&$+$2.48&$-$4.86 (0.35)& D&160&  0&19.3\\
 84780&V2113 Oph       &M5III     & 6.43& 6.43 (0.52)&$-$0.21&$-$6.18 (0.18)& D&372&  0&47.3 49.3 70.5 76.1 257.1\\
 84833&V656 Her        &M2III     & 5.01& 6.80 (0.36)&$+$0.89&$-$4.96 (0.13)& D&343&  0&22.3 93.9 176.1 234.7 283.3\\
 84938&NSV 21626       &M2.5III   & 6.35& 3.26 (0.49)&$+$2.08&$-$5.37 (0.33)& D&348&  0&19.7 23.6\\
 85435&V859 Ara        &M2II      & 7.07& 3.90 (0.58)&$+$1.94&$-$5.12 (0.40)&2M&372&  0&14.6 18.9 24.9 35.2 97.8 125.3\\
 85760&NO Aps          &M3III     & 5.83& 3.69 (0.28)&$+$0.75&$-$6.44 (0.17)& D&  0&207&26.2 26.6\\
 85934&V642 Her        &M4III     & 6.36& 5.42 (0.52)&$+$0.95&$-$5.40 (0.21)& D&364&  0&21.9 25.5 33.6 35.4 61.2 130.5 153.4 229.9\\
 86153&V959 Her        &M4III     & 6.37& 4.23 (0.44)&$+$1.41&$-$5.46 (0.24)& D&188&  0&25.3 35.7\\
 87390&HD 162189       &M2III     & 5.94& 5.77 (0.56)&$+$1.80&$-$4.40 (0.21)& G&207&  0&14.1 28.5\\
 88563&V980 Her        &M2III     & 6.97& 3.39 (0.53)&$+$2.27&$-$5.10 (0.42)&2M&358&  0&19.1 24.6 26.0 31.4 135.3\\
 88761&V983 Her        &M...      & 7.33& 3.28 (0.47)&$+$2.38&$-$5.05 (0.32)& D&  0&  0&\\
 89172&V669 Her        &M3III     & 4.96& 5.99 (0.22)&$+$0.33&$-$5.79 (0.08)& D&152&  0&22.9 24.0\\
 89527&V2392 Oph       &M4III     & 6.10& 4.22 (0.50)&$+$0.68&$-$6.22 (0.26)& D&379&  0&17.6 20.0 25.9 32.0 33.2 97.7\\
 91135&V4401 Sgr       &M4III     & 6.77& 6.46 (0.56)&$+$0.72&$-$5.24 (0.19)& G&387&  0&18.8 23.1 23.8 24.2 33.4 35.9\\
 91789&V536 Lyr        &M...      & 8.20& 2.64 (0.71)&$+$2.11&$-$5.81 (0.59)& D&127&  0&50.8 282.5\\
 92079&V4405 Sgr       &M4III     & 6.34& 4.49 (0.40)&$+$0.42&$-$6.33 (0.19)& G&291&  0&21.3 26.2 32.6 38.8 66.9\\
 93270&V387 Vul        &M3.5III   & 6.28& 4.63 (0.41)&$+$0.81&$-$5.89 (0.21)& G&294&  0&15.8 19.2\\
 93919&NSV 24691       &M1        & 7.53& 5.45 (0.72)&$+$1.93&$-$4.40 (0.29)& G&380&  0&12.0 15.5 103.2\\
 94646&HD 180377       &M2III     & 6.46& 3.88 (0.53)&$+$1.79&$-$5.29 (0.30)& G&331&  0&14.9 16.5 16.7 34.6 296.7\\
 94969&V389 Vul        &M4        & 6.95& 4.27 (0.57)&$+$0.39&$-$6.48 (0.29)& G&307&  0&36.1 49.4 58.9 82.3 226.8\\
 96111&V4419 Sgr       &M4III     & 6.58& 4.59 (0.62)&$+$1.03&$-$5.68 (0.30)& D&414&  0&29.3 39.9 57.4\\
 96310&V1919 Cyg       &M3III     & 6.64& 3.17 (0.39)&$+$1.50&$-$6.01 (0.27)& G& 15&  0&\\
 96399&NSV 24813       &M4/5III   & 6.71& 4.63 (0.59)&$+$1.21&$-$5.47 (0.28)& D&460&  0&11.9 12.0 13.9 15.3 21.4 29.0 135.1\\
 97142&V2090 Cyg       &M1III     & 6.39& 3.36 (0.36)&$+$0.74&$-$6.65 (0.24)& G& 11&  0&\\
 97365&$\delta$ Sge    &M2II+A0V  & 3.68& 5.49 (0.72)&$-$0.78&$-$7.09 (0.28)& D&286&  0&32.4 34.0 34.8 54.2\\
 98438&VZ Sge          &M4III     & 5.33& 3.20 (0.31)&$-$0.10&$-$7.59 (0.21)& G&319&  0&36.5 39.2 51.4 65.1\\
 98500&V345 Tel        &M4III     & 6.75& 5.40 (0.59)&$+$1.50&$-$4.85 (0.24)& G&461&  0&17.2 20.5 22.4 42.3 75.4\\
 98608&NU Pav          &M6III     & 4.95& 6.86 (0.26)&$-$1.47&$-$7.29 (0.09)& D&434&112&59.0 62.0 87.0 124.7 272.5\\
 98688&V3872 Sgr       &M4III     & 4.43& 7.27 (0.18)&$-$0.76&$-$6.46 (0.06)& D&367&  0&24.0 30.4 31.3 42.8 50.5 234.7\\
 98954&V1472 Aql       &M2.5III   & 6.37& 7.92 (1.07)&$+$1.73&$-$3.78 (0.29)& G&318&  0&100.2\\
100051&V2114 Cyg       &M5        & 7.23& 2.50 (0.51)&$+$0.86&$-$7.18 (0.44)& G& 12&  0&\\
101316&MT Del          &M4II-III  & 7.12& 3.39 (0.66)&$+$2.13&$-$5.23 (0.42)& G&307&  0&26.0 29.0 50.7 64.1 103.7\\
101810&EU Del          &M6III     & 6.22& 8.56 (0.50)&$-$1.12&$-$6.46 (0.14)& G&265&  0&44.0 60.8 67.3 132.6 235.3\\
102096&AV Mic          &M3II      & 6.28& 2.85 (0.40)&$+$1.10&$-$6.64 (0.31)& D&438&  0&22.3 23.3 30.3 31.0 32.3 45.0 110.7\\
102624&EN Aqr          &M3III     & 4.43& 5.57 (0.28)&$-$0.22&$-$6.50 (0.11)& D&346&  0&20.2 24.9 27.2 35.0 36.9 143.9 197.2\\
102720&FI Vul          &M3        & 7.65& 5.15 (0.60)&$+$0.16&$-$6.29 (0.25)& D&270&  0&53.8 64.6 73.7 111.7\\
103645&V2142 Cyg       &M3III     & 6.85& 3.53 (0.43)&$+$2.47&$-$4.81 (0.30)& D& 88&  0&22.3\\
104755&NSV 25504       &M1-2III   & 5.06& 3.65 (0.67)&$+$0.69&$-$6.51 (0.40)& D&424&  0&15.5 88.3 130.9 162.6 228.3\\
104974&NSV 13620       &M3III     & 5.31& 5.33 (0.33)&$+$0.54&$-$5.84 (0.14)& D&348&  0&19.5 20.1 24.3 91.5\\
105334&T Ind           &C7,2(Na)  & 6.15& 1.72 (0.40)&$+$0.60&$-$8.22 (0.51)& D&  0& 51&157.5\\
105518&NSV 25556       &M2III     & 7.10& 2.93 (0.57)&$+$2.81&$-$4.87 (0.49)&2M&348&  0&16.6\\
105678&Y Pav           &C7,3(N0)  & 6.28& 2.48 (0.59)&$+$0.33&$-$7.70 (0.52)& D&  0&152&232.0\\
105846&V2158 Cyg       &M...      & 7.77& 4.18 (0.59)&$+$1.99&$-$4.92 (0.33)& D&  0&  0&\\
106044&SX Pav          &M5III     & 5.47& 7.92 (0.28)&$-$0.71&$-$6.22 (0.34)&2M&422&236&36.0 50.6 53.1 77.6 88.9\\
106062&NV Peg          &M4.5III   & 5.84& 4.90 (0.32)&$-$0.16&$-$6.72 (0.14)& D&261&  0&44.0 45.4 77.0 91.5\\
107516&EP Aqr          &M8III     & 6.52& 8.80 (0.63)&$-$1.53&$-$6.81 (0.16)& D&324&  0&81.8 117.9 125.8\\
109212&OY Peg          &M1        & 6.31& 3.26 (0.38)&$+$1.62&$-$5.83 (0.26)& D&260&  0&28.1 29.4 38.6 41.0 64.3 202.0\\
110256&BO Oct          &M5III     & 5.09&11.22 (0.23)&$-$1.16&$-$5.92 (0.05)& D&  0&315&18.0 23.1 23.4 201.6\\
110346&PT Peg          &M4III     & 6.46& 4.43 (0.48)&$+$1.35&$-$5.43 (0.24)& D&163&  0&164.7\\
110364&DN Oct          &M3III     & 7.30& 2.89 (0.47)&$+$2.06&$-$5.67 (0.36)& D&232&  0&17.9 32.4\\
110428&BW Oct          &M7III     & 8.73& 4.00 (0.66)&$-$0.27&$-$7.29 (0.36)& D&  0&  0&\\
110478&$\pi^1$ Gru     &S5        & 6.42& 6.13 (0.76)&$-$2.15&$-$8.21 (0.27)& D&  0&101&198.8\\
111043&$\delta^2$ Gru  &M4.5III   & 4.12& 9.88 (0.23)&$-$0.87&$-$5.90 (0.06)& G&370&212&20.6 24.1 24.5 32.3 33.3 97.6 125.9 205.8\\
111310&$\nu$ Tuc       &M4III     & 4.91&11.24 (0.23)&$-$0.14&$-$4.89 (0.05)& D&408&210&22.3 24.4 24.8 25.1 25.5 33.8 50.6 80.1 123.2 261.8\\
112102&NSV 25945       &M5III     & 6.12& 4.73 (0.39)&$+$1.13&$-$5.50 (0.18)& D&371&  0&13.2 15.5 17.2 27.4 106.3\\
112122&$\beta$ Gru     &M5III     & 2.07&18.43 (0.42)&$-$3.21&$-$6.88 (0.05)& D&303&245&28.1 37.1 37.7 38.4 39.2 39.7 105.6 232.6\\
112961&$\lambda$ Aqr   &M2.5III   & 3.73& 8.47 (0.66)&$-$0.64&$-$6.01 (0.17)& D&273&  0&24.5 32.0 49.5\\
113330&DM Tuc          &M8III     & 7.43& 4.49 (0.50)&$-$0.09&$-$6.83 (0.24)& D&430&  0&137.6 153.6 170.4 208.3\\
113881&$\beta$ Peg     &M2.5II-III& 2.44&16.64 (0.15)&$-$2.21&$-$6.11 (0.07)& G&115&  0&43.3\\
114318&Y Scl           &M4        & 8.43& 6.05 (0.81)&$+$0.39&$-$5.71 (0.30)& D&377&  0&95.7 291.5\\
114347&GZ Peg          &M4III     & 5.05& 4.17 (0.34)&$-$0.45&$-$7.36 (0.18)& D&273&  0&40.1 42.2 46.5 49.4 80.3\\
114404&V345 Peg        &M3        & 6.69& 4.01 (0.51)&$-$0.06&$-$7.06 (0.28)& D&  8&  0&\\
114407&DL Gru          &M4III     & 5.90& 3.55 (0.43)&$+$0.74&$-$6.51 (0.26)& D&398&  0&18.6 23.4 24.0 24.5 29.5 30.1 162.1\\
114939&$\chi$ Aqr      &M3III     & 4.93& 5.32 (0.37)&$-$0.29&$-$6.67 (0.15)& D&299&  0&32.3 38.5 44.9\\
115433&DR Tuc          &M3III     & 6.08& 3.84 (0.44)&$+$1.37&$-$5.72 (0.25)& D&452& 52&24.0 33.0 47.4 254.5\\
116264&HW Peg          &M5III     & 5.33& 6.27 (0.28)&$-$0.18&$-$6.21 (0.10)& D&213&  0&30.7 41.7 43.2\\
116307&NSV 26103       &M3III     & 6.10& 3.60 (0.35)&$+$1.04&$-$6.19 (0.21)& D&224&  0&34.7 102.7\\
117276&CK Phe          &M2III     & 7.30& 3.31 (0.64)&$+$2.57&$-$4.84 (0.43)& D&410&  0&21.8 22.4 29.3 42.8 137.4\\
117628&HH Peg          &M3III     & 5.77& 3.86 (0.43)&$+$1.25&$-$5.85 (0.24)& G&254&  0&21.1 40.2 113.5\\
117887&XZ Psc          &M5III     & 5.78& 5.12 (0.41)&$+$0.13&$-$6.33 (0.18)& D&261&  0&30.2 39.4 41.2\\
117986&V363 Peg        &M3.5III   & 6.45& 4.93 (0.40)&$+$1.27&$-$5.27 (0.18)& D&241&  0&21.9 22.7 32.3 33.0\\
118131&$\psi$ Peg      &M3III     & 4.63& 6.85 (0.24)&$+$0.03&$-$5.81 (0.09)& D&169&  0&37.4 118.9\\

\end{longtable}
\twocolumn
\end{center}

\end{scriptsize}

%%%%%%%%%%%%%%%%%%%%%%%%%%%%%%%%%%%%%%%%%% end multi-page %%%%%%%%%%%%%%%%%

\subsection{Comparison with known periods}

\begin{figure}
 \includegraphics[scale=1.0, angle=0]{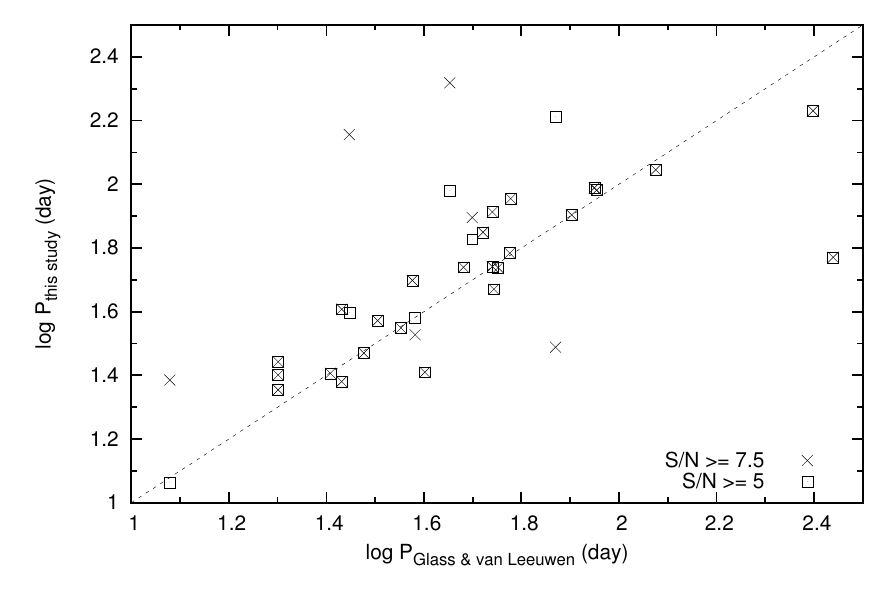}
 \caption{A comparison of periods for stars appearing in \citet[][]{b_gla07} and this study. Periods are shown for two S/N limits. Overlapping symbols indicate that no closer period was found by relaxing the S/N limit.}
 \label{fig012}
\end{figure}

We compared our derived periods to those found in the literature and found that 21 stars in our sample are listed in \citet{b_gla07}. Using each listed period less than 300 d, we searched for the closest corresponding period from our study. Figure~\ref{fig012} shows a comparison using periods selected with two different S/N limits. Although the agreement is quite good with a S/N limit of 7.5, reducing the limit to S/N $\ge$ 5 found matching periods for some lower luminosity stars, which tend to have smaller amplitudes. The agreement is remarkably good considering that SRVs show long-term amplitude modulation or mode-switching \citep{b_bed98,b_kis99}. \citet{b_gro04} compared the periods of 370 stars over a time span of 17 years and found that 30 stars showed $> 10\%$ change in period, while an additional 36 stars exhibited period changes large enough to require assignment to a different P-L sequence.

% 091/x.eps

\subsection{Variability and multi-periodicity}

\begin{figure}
 \includegraphics[scale=1.0, angle=0]{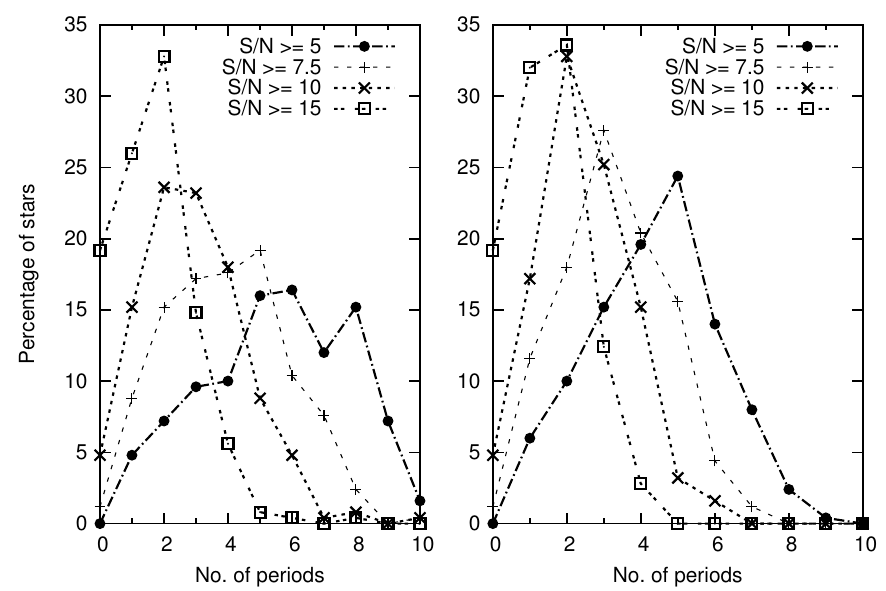}
 \caption{The number of periods per star as a percentage of the sample, plotted for S/N limits of 5, 7.5, 10, and 15. The left panel shows all periods, while in the right panel, closely spaced periods  have been removed, as explained in the text.}
 \label{fig013}
\end{figure}

In their recent review article, \citet{b_eye08} stated that ``\emph{all} red giants with spectral types later that early K are variable'', which led us to investigate whether we could confirm this within our sample.
Excluding 11 stars that lacked sufficient observations for period determination, we find that all stars are variable at a detection threshold of S/N $\ge 5$. Figure~\ref{fig013} shows the number of periods detected as a percentage of the effective sample size (250) for a range of S/N thresholds. The number of stars with no detected period was 0, 3, 12, 48 for S/N limits of 5, 7.5 (used for Table \ref{T4}), 10, and 15, respectively. Since we selected the majority of our sample using the Hipparcos catalogue's course variability flag (H6; section \ref{s_samp}), our results confirm it is a reliable predictor of variability. However, several stars
exhibited clear variability but with amplitudes $<$ 0.01 mag, far less than the amplitude observed by Hipparcos (H6 $\ge$ 2 indicates observed amplitudes $\ge$ 0.06 mag), although it should be noted that the amplitudes from iterative sinewave fitting are smaller than the peak-to-peak amplitudes of each time-series. We defer further discussion of amplitudes to Paper II.

It is well known that semi-regular variables in the Milky Way exhibit multi-periodic behaviour. \citet{b_mat97} analysed long-term visual light curves of SRVs using data from the AAVSO database, concluding that most SRVs are multi-periodic, and exhibit period ratios $1.7 \le P_1 / P_2 \le 1.95$. They also confirmed 6 SRVs to be triply-periodic. An analysis of long-term visual observations of 93 red SRVs by \citet{b_kis99} included only the most significant periods and found mono-periodic behaviour in 31\% of the sample, two significant periods for 47\%, with a further 13\% exhibiting three dominant periods. By contrast, we have extracted all periods above specific S/N thresholds, and include closely spaced periods. In order to more easily compare our results with earlier studies, we sorted periods for each star into ascending order, and omitted period $P_{\rm i}$ when $P_{\rm i+1}$ / $P_{\rm i} <$ 1.1, since they are likely to represent the same mode (Figure \ref{fig013}, right panel).

Figure~\ref{fig013} shows that the modal number of periods for S/N limits of 10 and 15 is 2, which agrees with the earlier studies. Likewise, our distributions are remarkably similar to that of \citet{b_kis99}, with the primary difference being that, as the S/N constraint is lowered, a small number of additional periods have been identified. While the curves with S/N $\ge$ 10 are quite similar in both plots, the S/N $\ge$ 5 and S/N $\ge$ 7.5 curves in the left panel are broader and not quite as smooth as those in the right panel, indicating that as the S/N threshold is lowered, additional closely spaced frequencies (of lower amplitude) are identified.

If both frequencies were of similar amplitude, simultaneously present, and with long mode lifetimes, we would expect to see the effect of beating in the light curves, which we do not. Although beyond the scope of the current paper, we briefly note that we analysed several stars with close periods and well sampled LCs, consisting of both CCD and PEP observations. By splitting the datasets (temporally) into halves, and using both Fourier and wavelet analysis, we confirmed that the dominant frequencies and amplitudes changed between the two segments. The results were consistent for separate analyses of PEP-only, and CCD-only subsets, as well as the combined dataset. Thus it appears that amplitude and/or phase modulation is the likely cause of the closely spaced periods, consistent with stochastically excited oscillations.

We note that previous analyses of OGLE and MACHO photometry of the Magellanic Clouds have used S/N limits between 3 and 10 \citep[see, for example,][]{b_sos04,b_der}.

% 093/x.eps

\subsection{Period ratios}

% Fig 3.19 ,  81/v2
\begin{figure}
 \includegraphics[scale=1.0, angle=0]{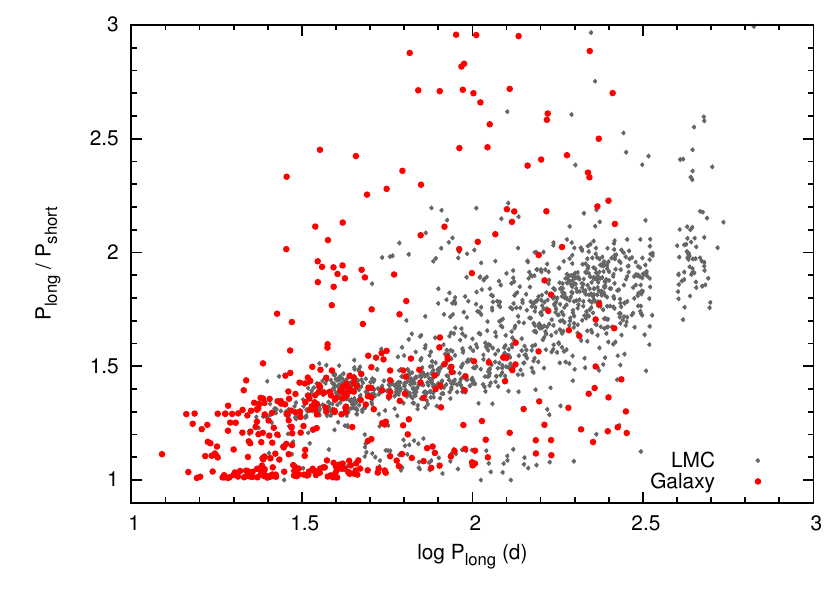}
 \caption{Petersen diagram plotting $\log P_{\rm long}$ against the ratio of $P_{\rm long}$ to $P_{\rm short}$ for adjacent periods (red) in galactic M giants (S/N $\ge 10$). Data for the LMC \citep{b_der} are plotted in grey.}
 \label{fig014}
\end{figure}

Period ratios provide a useful mechanism to compare the pulsational behaviour of local SRVs to those in the lower metallicity environment of the LMC, since they are independent of distance. Figure~\ref{fig014} plots a Petersen diagram for all multi-periodic stars in our sample (S/N $\ge 10$). Each red point represents the ratio of a longer period to the shorter, plotted against the log of the longer period. For comparison, we have plotted the period ratios of LMC MACHO red giants \citep[][grey points]{b_der}, which show a striking similarity. The most densely populated region occurs at a period ratio of about 1.3, corresponding to first and second radial overtone modes ($P_1/P_2$) in the theoretical model of \citet{b_woo96}. The ratio is in excellent agreement with the MACHO data, and is one of the most common period ratios found by \citet{b_sos04} in their analysis of period ratios in OGLE observations of the LMC.

The other significant overlap occurs for period ratios close to 1. Although the MACHO data contain few short-period ($\log P < 1.4$) stars in this region, the analysis of OGLE data by \citet{b_sos04} appears to have identified many more variables, and their Petersen diagram (fig 11) shows a large number of stars extending to $\log P \sim 1.1$, in agreement with our local SRVs. A small period ratio has many possible interpretations, including two closely separated oscillation modes, high-overtone radial pulsation, a single, slightly varying period, or may be an artefact of amplitude or phase modulation. The former interpretation may imply the presence of non-radial oscillation modes \citep{b_kis06}. However, as noted earlier, our data include multiple closely spaced periods that probably represent the same oscillation mode, which will result in the small period-ratios being over represented in our Petersen diagram, although this is somewhat reduced by the use of periods with S/N $\ge$ 10.

The area of minimal overlap with LMC data around $\log P \sim$ 2.4 and period ratios of $\sim$ 1.9 is most likely due to our sample selection, which excluded stars with GCVS periods longer than 100 d. Although they have no obvious analogue in the MACHO data, we note that the small group of stars with period ratios of $\sim$ 1.9 (and $\log\;P \sim 1.6$) do appear in the larger LMC sample of \citet{b_sos04}. Furthermore, our period ratios are consistent with those found by \citet{b_wra}, who demonstrated that ratios of $\sim$ 1.0, 1.1, 1.3, and 1.9 are common amongst pulsating red giants in the galactic bulge.

\section{Conclusion \& future work}
\label{s_conclusion}

A fully autonomous instrument has been designed, built, and operated for 5.5 years for a unique, homogenous survey of the nearest M giant stars. The main results of this paper are the following:
\begin{enumerate}
\item We have determined the characteristic pulsation periods of 247 bright, nearby, M giants, the majority of which have not been determined previously, thereby increasing the number of such stars by a factor of 4.
\item We provide updated $K$-band magnitudes, selected from several NIR sources, which we carefully compared to determine the most reliable values. We include the effect of interstellar extinction and show that circumstellar extinction can be ignored.
\item We employ a novel technique to correct residual errors in our flat-fielding images, and successfully use Optimal Image Subtraction for a subset of blended stars, yielding scintillation-limited differential photometry to better than 1\% precision.
\item Twenty-three stars were observed concurrently with PEP and are in excellent agreement with the CCD data. Moreover, 21 stars in our sample had periods determined previously, and a comparison with literature values showed our periods to be in good agreement.
\item All stars in the sample were found to be variable, leading us to conclude that the Hipparcos variability flag, which was used to select the sample, is a good predictor of variability.
\item Many stars in the sample exhibit closely spaced periods (period ratios $<$ 1.1), similar to those seen in solar-like stars, which probably represent changes in amplitude and/or phase of a single oscillation mode.
\item Eighty-seven percent of the stars were found to be multi-periodic, with the modal number of periods being 2 (S/N $\ge$ 10) and 3 (S/N $\ge$ 7.5).
\item The period ratios of local M giants were compared to multi-periodic stars in the LMC and Bulge, and were found to be in excellent agreement.
\end{enumerate}
Our motivation for conducting this survey was to significantly increase the sample of nearby M giants to enable unambiguous identification of the local P-L sequences, and to determine the metallicity dependence of the P-L zero-point. These issues will be discussed in Paper II.

\section*{Acknowledgments}

We extend our thanks to Aliz Derekas for kindly supplying her LMC data for Figure \ref{fig014}. This research has made use of the VizieR and SIMBAD databases, operated at CDS, Strasbourg, France. This project has been supported by the Australian Research Council and the Hungarian OTKA Grant K76816.

\appendix
\section{Light curves}

\begin{comment} #######################
\begin{figure*}
 \includegraphics[scale=1.0, angle=0]{fig019_224}   % 013064
 \caption{A sample light curve. The full appendix is available in the online version of this paper. CCD and PEP observations are denoted by red(solid) and blue(open) points, respectively. Hipparcos catalogue numbers are listed in parentheses. }
 \label{fig019}
\end{figure*}

\begin{figure*}
 \includegraphics[scale=1.0, angle=0]{fig019_224p}
 \caption{Power spectrum corresponding to the light curve in Figure~\ref{fig019}. The inset shows the spectral window using the same horizontal scale as the main plot. A horizontal line shows the mean signal in the frequency range 0.20--0.40 cycle d$^{-1}$. Peaks corresponding to periods listed in Table \ref{T4} are indicated with dotted lines.}
 \label{fig019p}
\end{figure*}
\end{comment} #######################

% 088
\begin{figure*}
 \includegraphics[scale=1.0, angle=0]{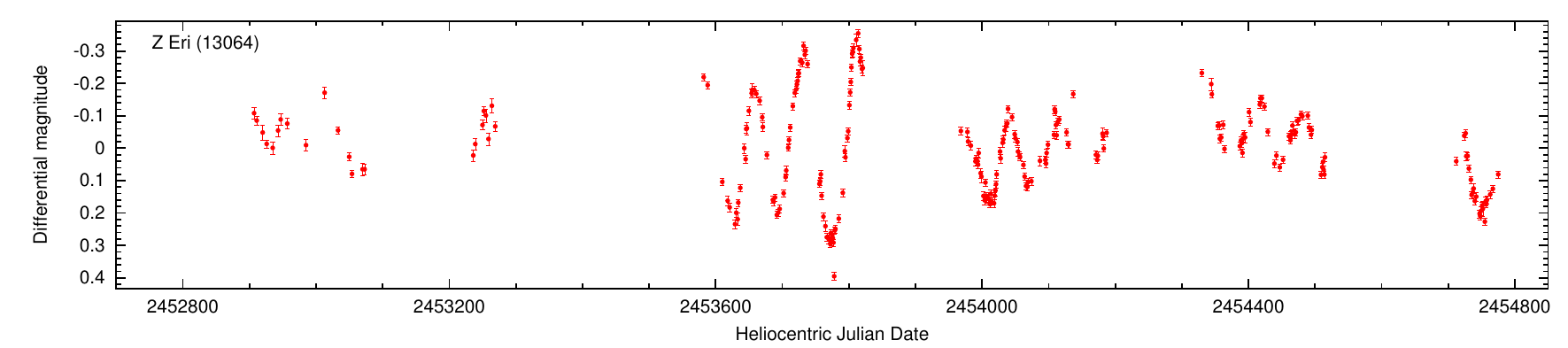}   % 013064
 \includegraphics[scale=1.0, angle=0]{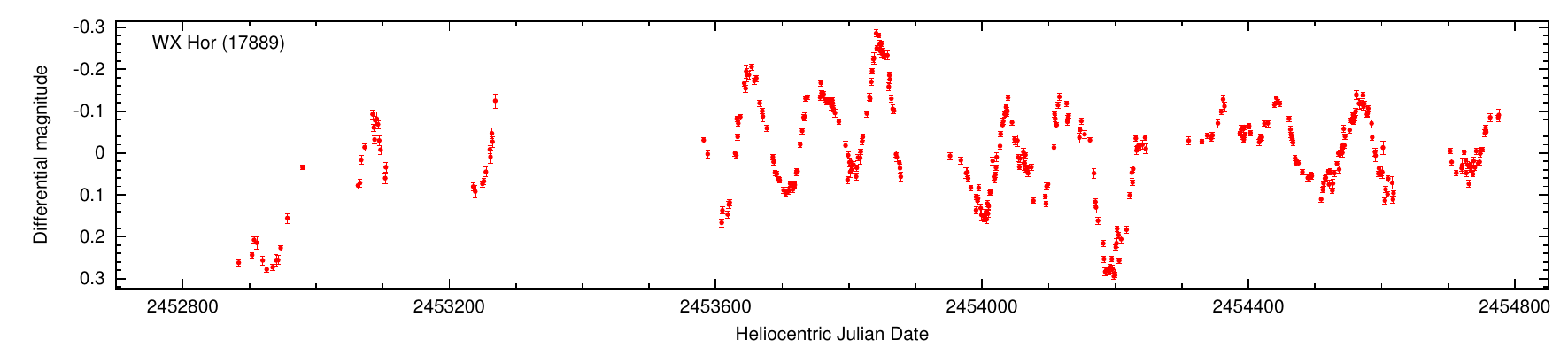}   % 017889
 \includegraphics[scale=1.0, angle=0]{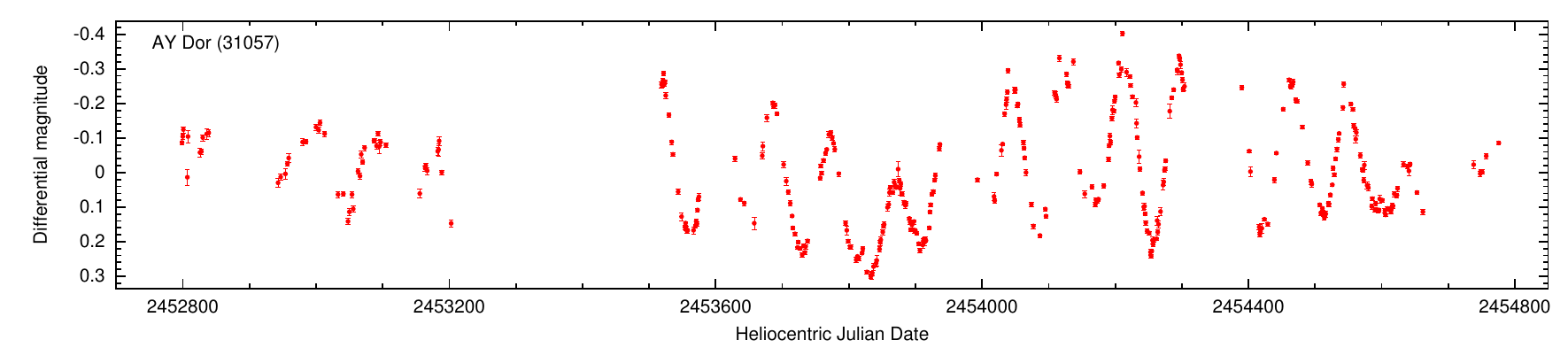}   % 031057
 \includegraphics[scale=1.0, angle=0]{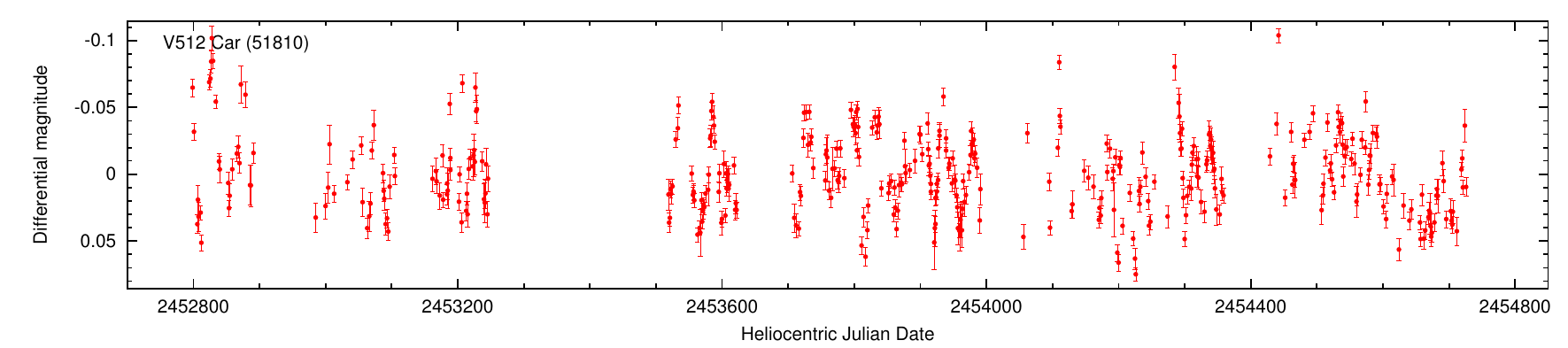}   % 051810
 \includegraphics[scale=1.0, angle=0]{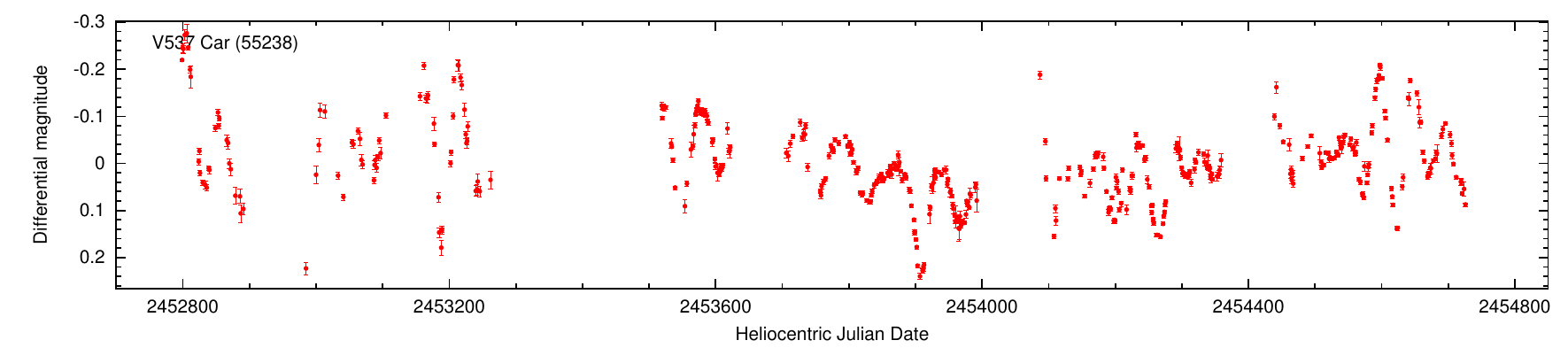}   % 055238
 \caption{A selection of light curves. CCD and PEP observations are denoted by red(solid) and blue(open) points, respectively. Hipparcos catalogue numbers are listed in parentheses.}
 \label{fig019}
\end{figure*}

\begin{figure*}
 \includegraphics[scale=1.0, angle=0]{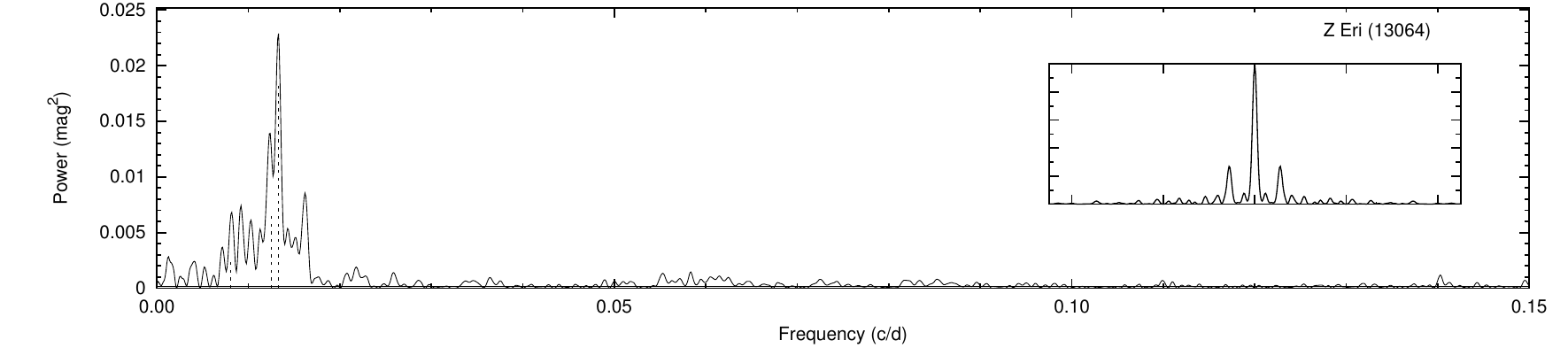}
 \includegraphics[scale=1.0, angle=0]{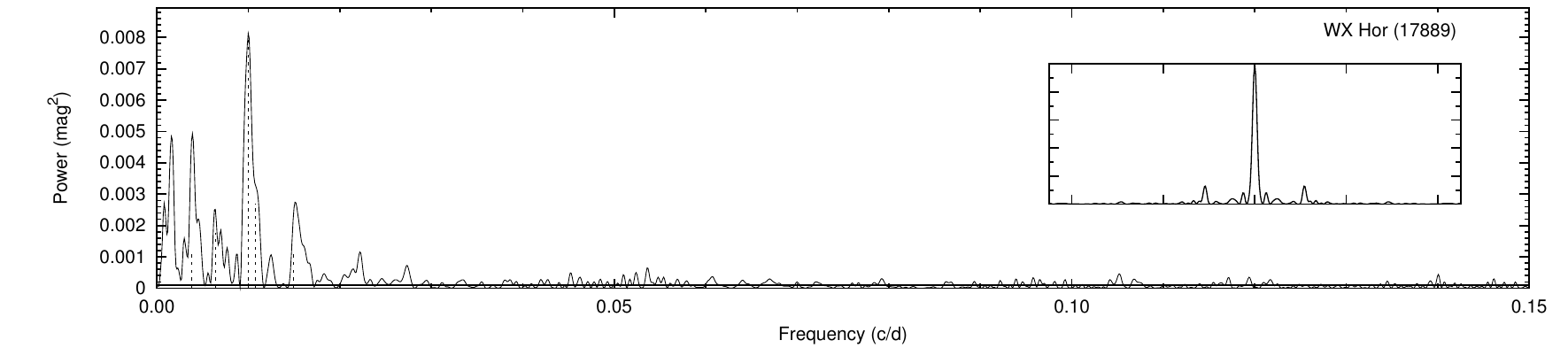}
 \includegraphics[scale=1.0, angle=0]{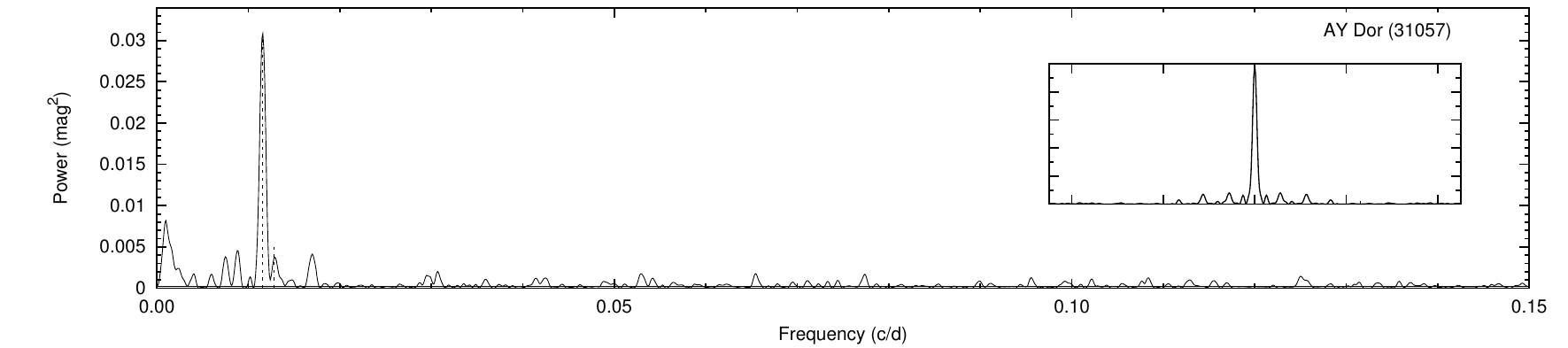}
 \includegraphics[scale=1.0, angle=0]{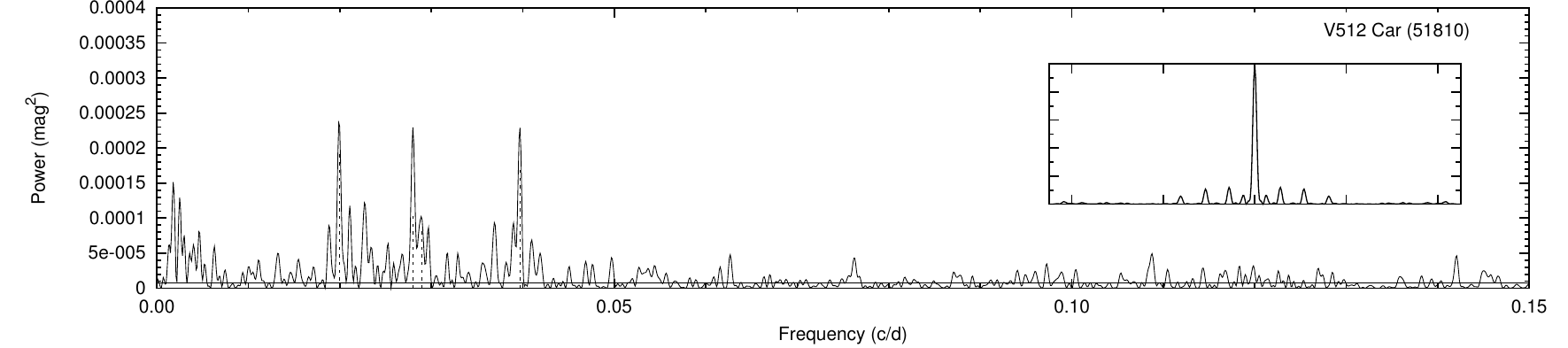}
 \includegraphics[scale=1.0, angle=0]{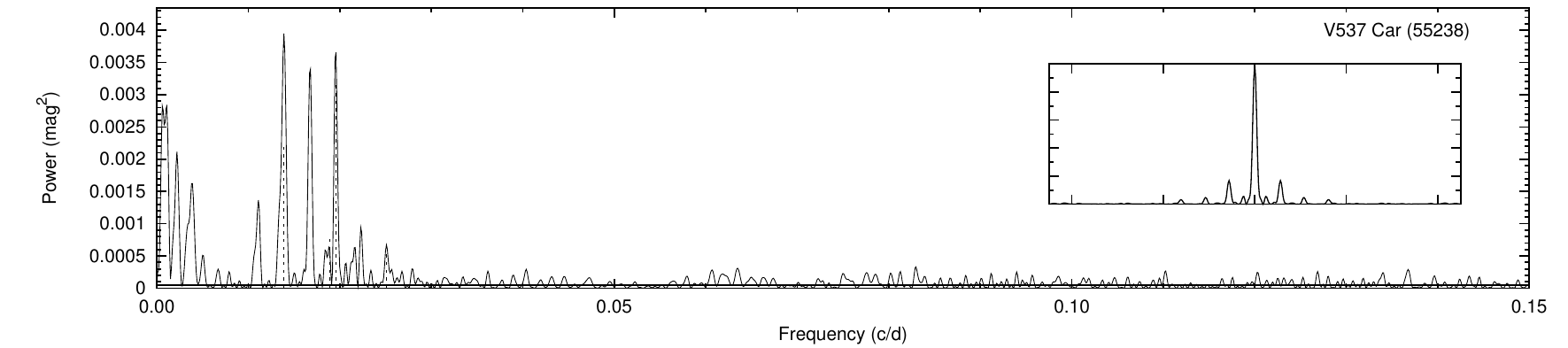}
 \caption{Power spectra corresponding to the light curves in Figure~\ref{fig019}. Insets show the spectral window using the same horizontal scale as the main plot. A horizontal line shows the mean signal in the frequency range 0.20--0.40 cycle d$^{-1}$. Peaks corresponding to periods listed in Table \ref{T4} are indicated with dotted lines.}
 \label{fig019p}
\end{figure*}

\begin{figure*}
 \includegraphics[scale=1.0, angle=0]{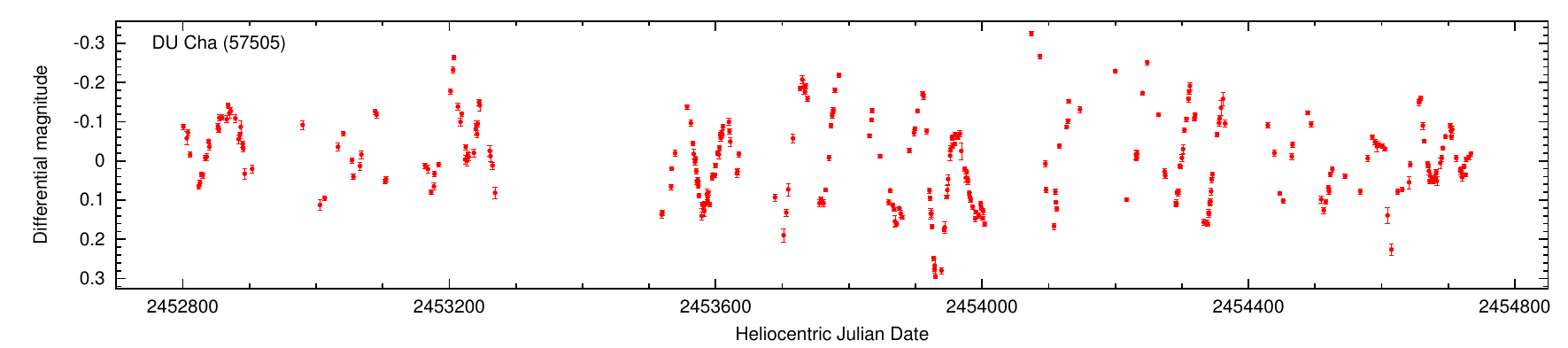}   % 057505
 \includegraphics[scale=1.0, angle=0]{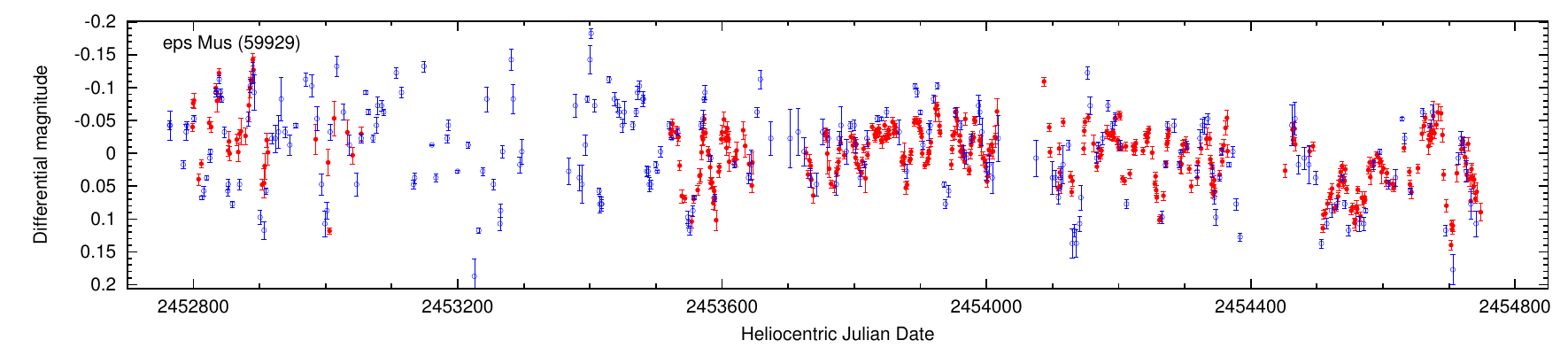}   % 059929
 \includegraphics[scale=1.0, angle=0]{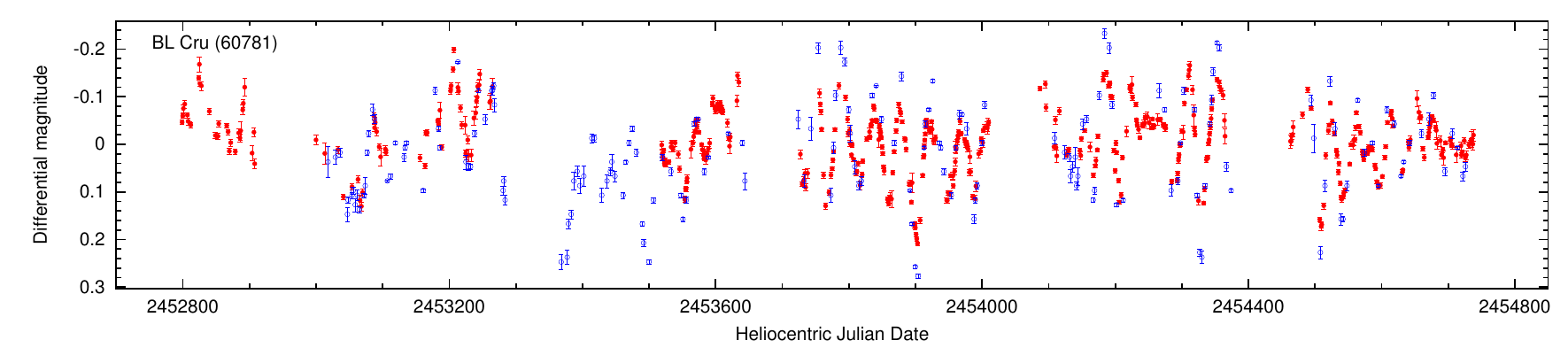}   % 060781
 \includegraphics[scale=1.0, angle=0]{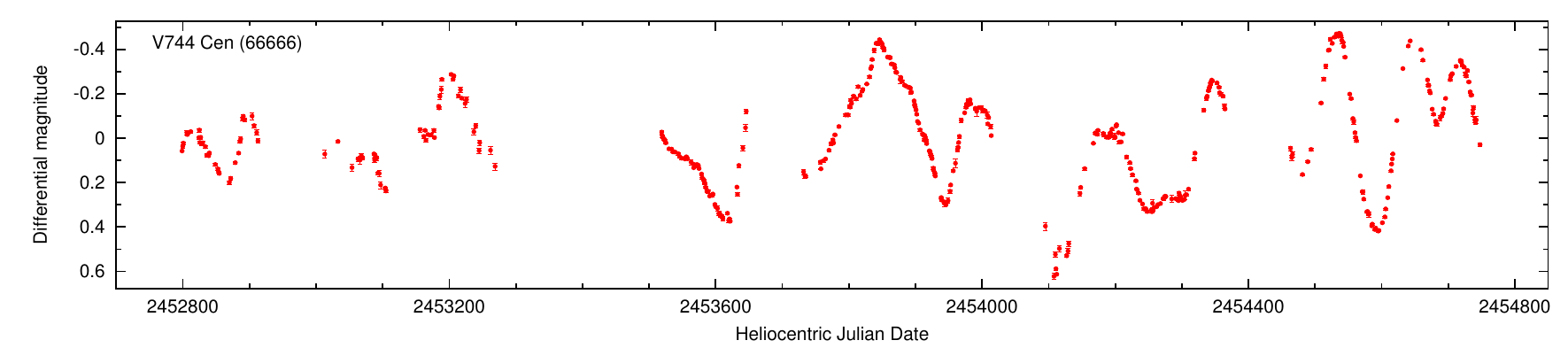}   % 066666
 \includegraphics[scale=1.0, angle=0]{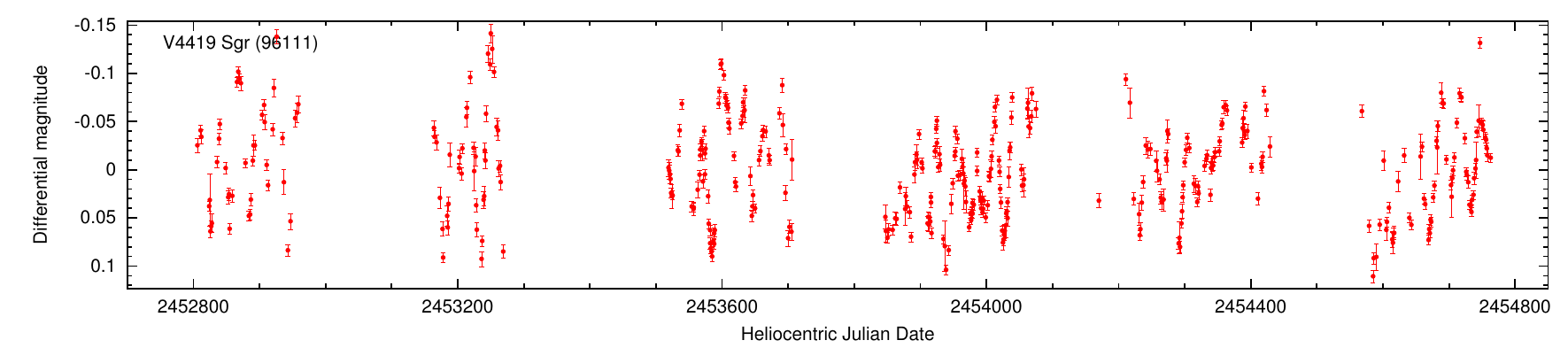}   % 096111
 \caption{A selection of light curves.}
 \label{fig020}
\end{figure*}

\begin{figure*}
 \includegraphics[scale=1.0, angle=0]{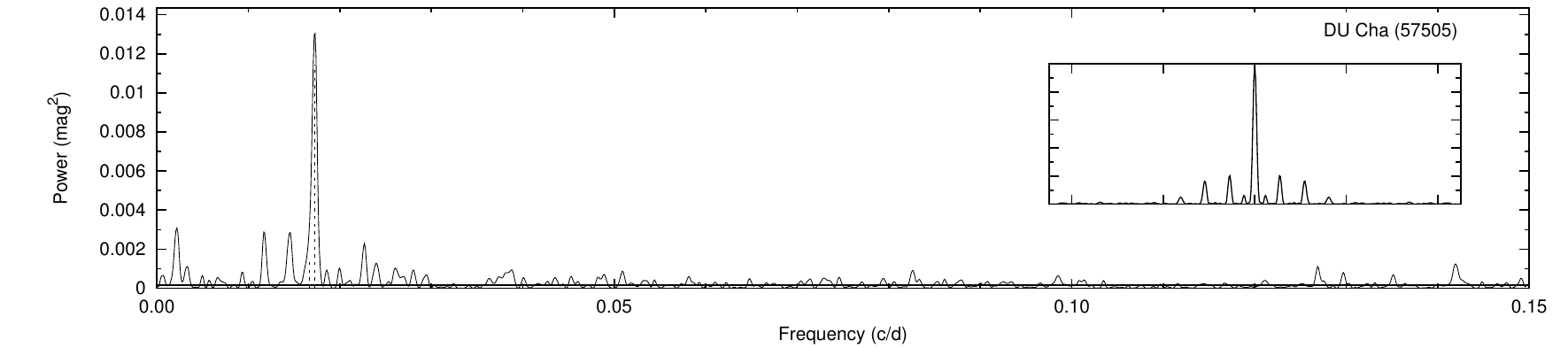}
 \includegraphics[scale=1.0, angle=0]{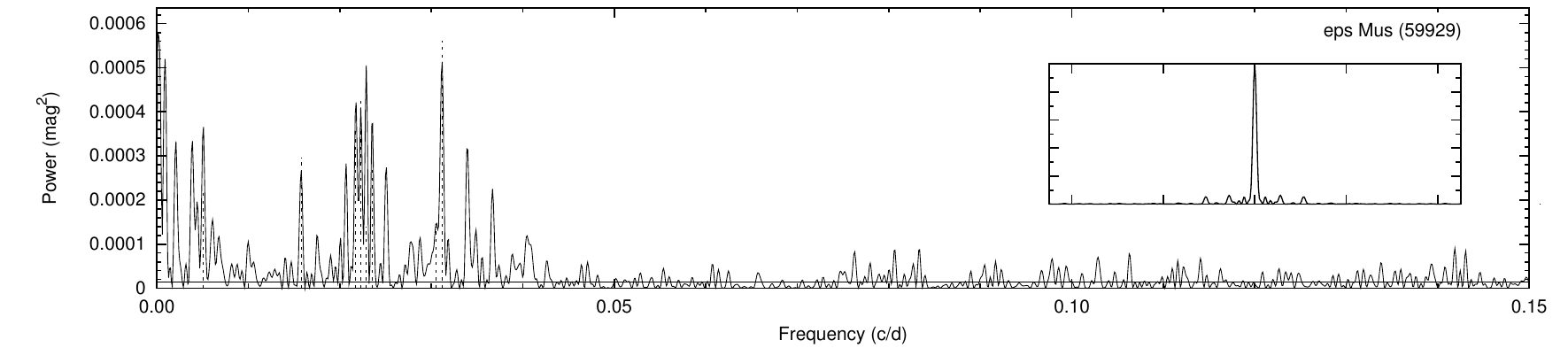}
 \includegraphics[scale=1.0, angle=0]{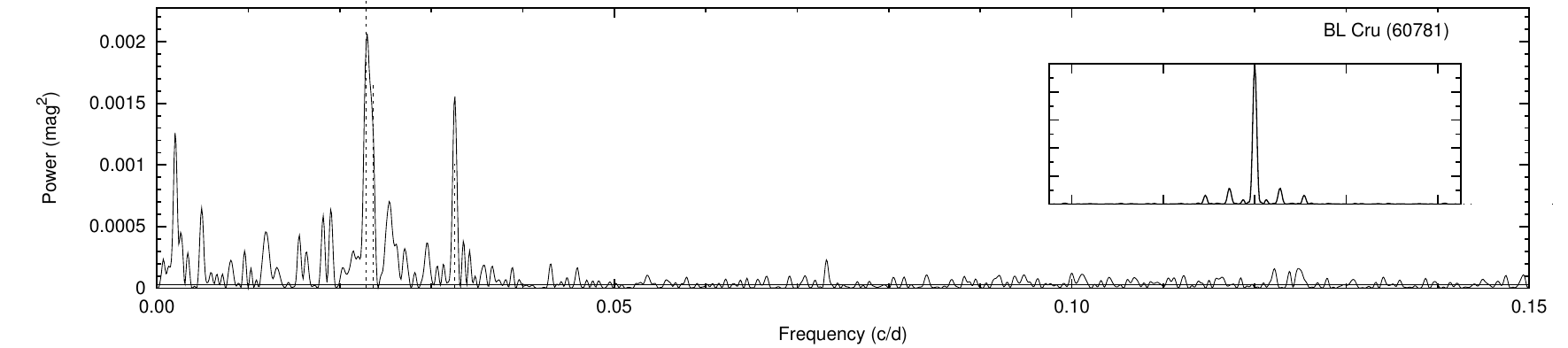}
 \includegraphics[scale=1.0, angle=0]{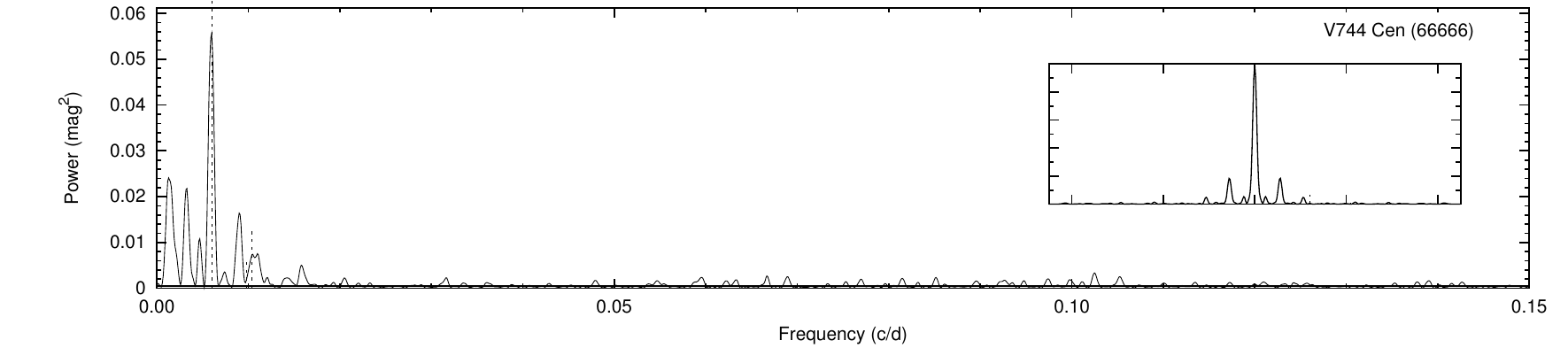}
 \includegraphics[scale=1.0, angle=0]{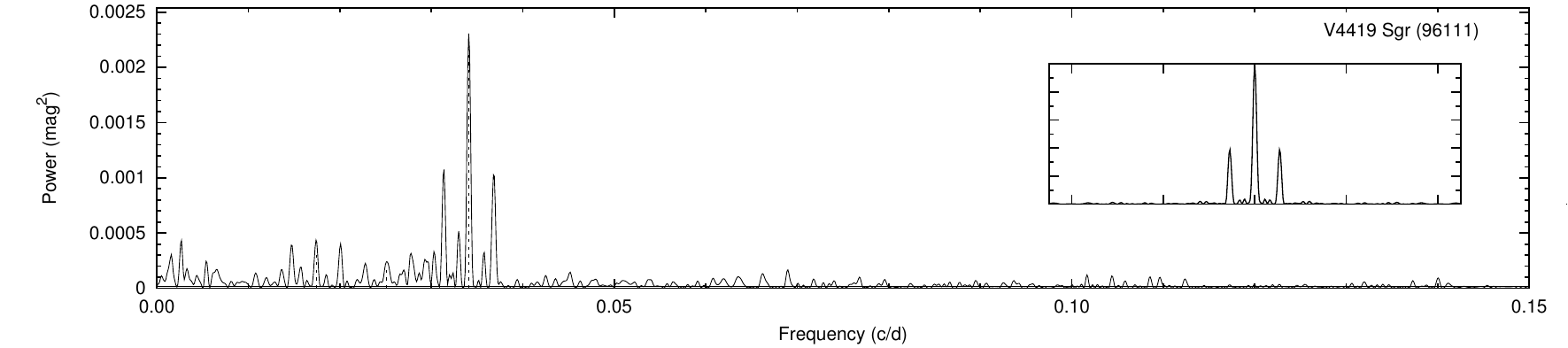}
 \caption{Power spectra corresponding to the light curves in Figure~\ref{fig020}.}
 \label{fig020p}
\end{figure*}

\begin{figure*}
 \includegraphics[scale=1.0, angle=0]{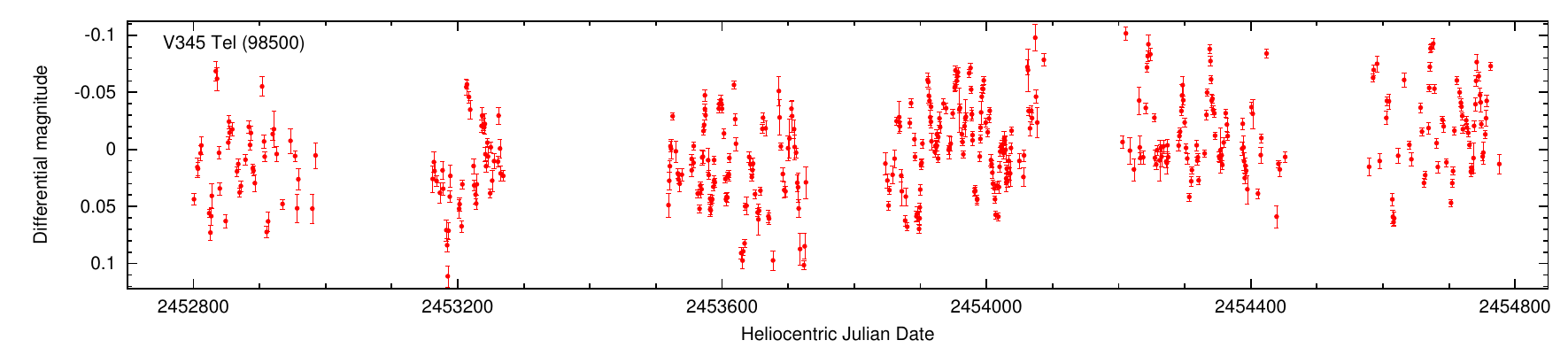}   % 098500
 \includegraphics[scale=1.0, angle=0]{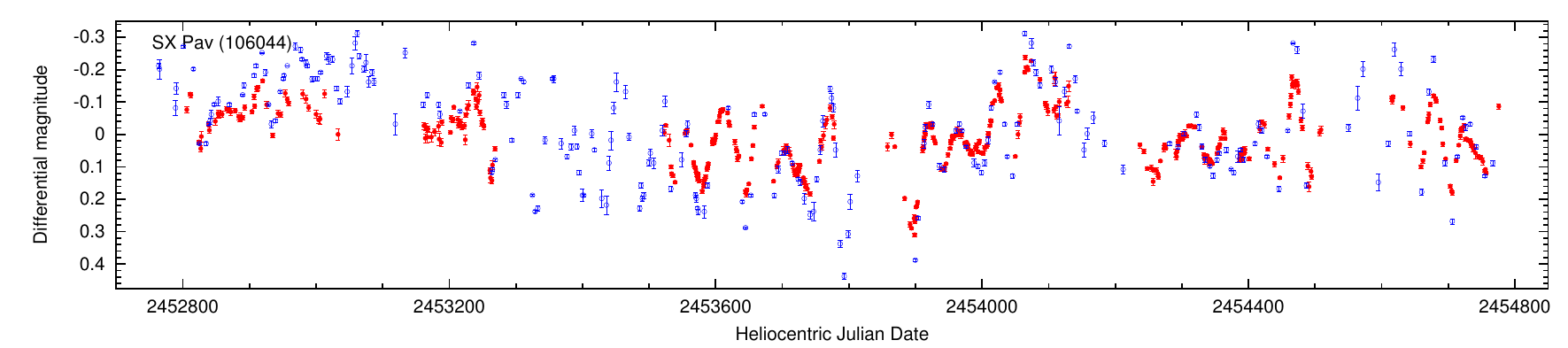}   % 106044
 \includegraphics[scale=1.0, angle=0]{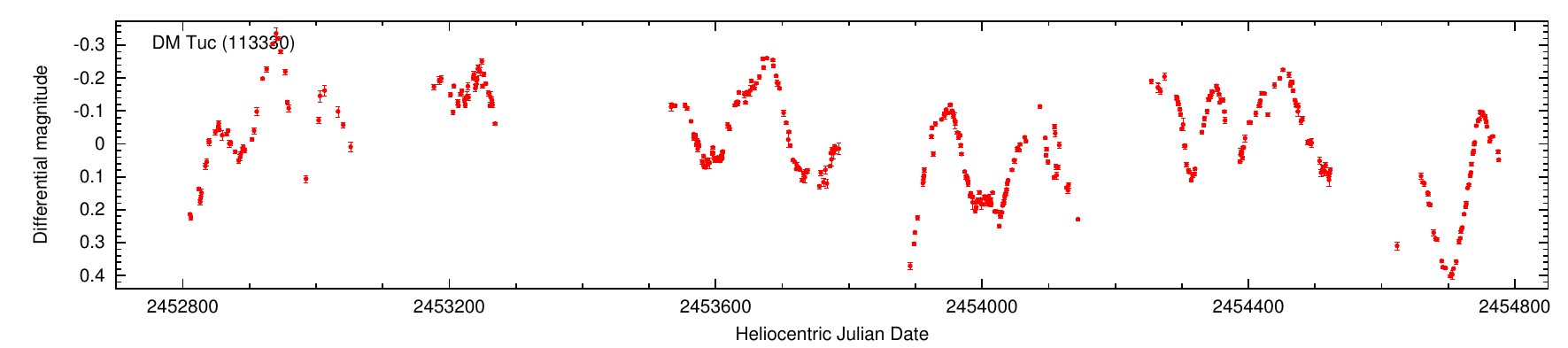}   % 113330

 \includegraphics[scale=1.0, angle=0]{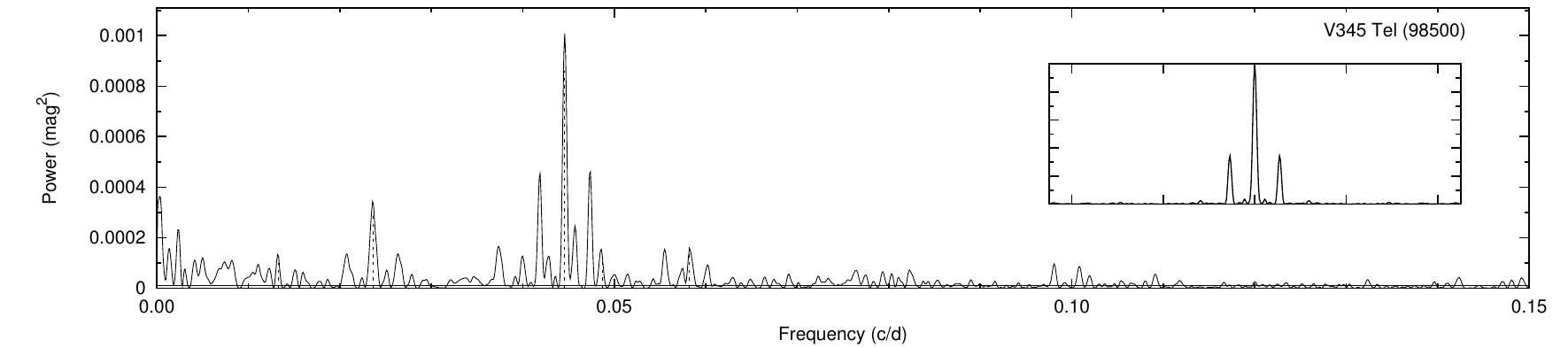}
 \includegraphics[scale=1.0, angle=0]{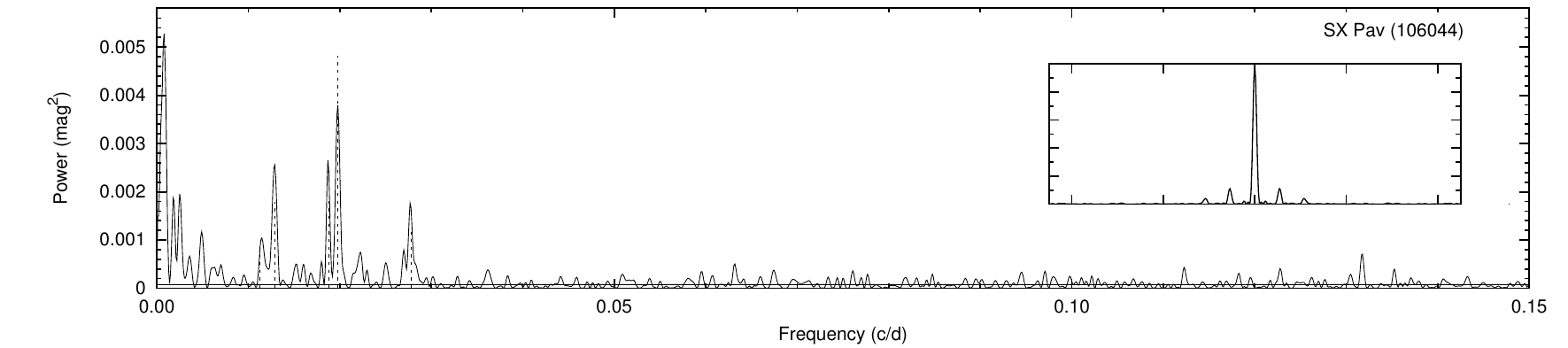}
 \includegraphics[scale=1.0, angle=0]{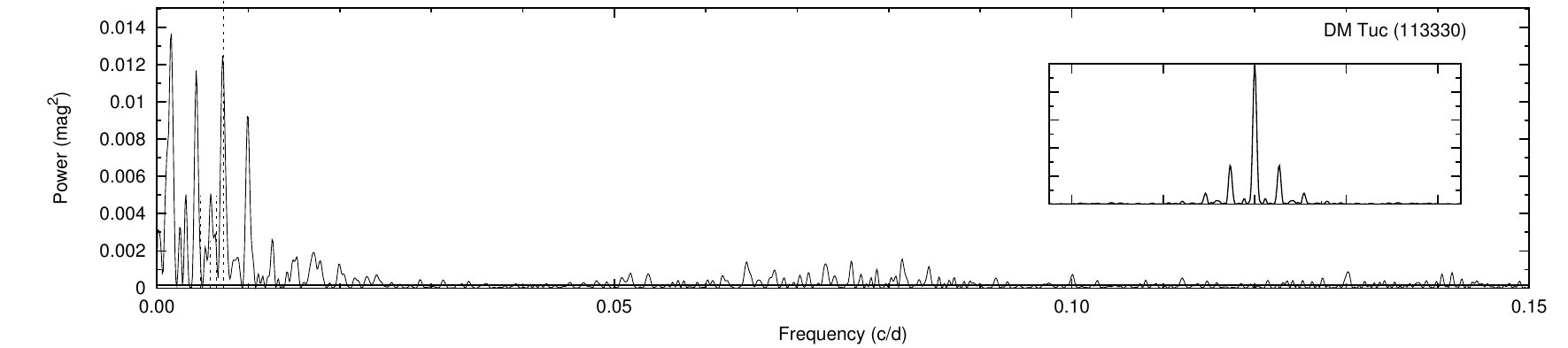}
 \caption{A selection of light curves and corresponding power spectra.}
 \label{fig022}
\end{figure*}

\begin{figure*}
 \includegraphics[scale=1.0, angle=0]{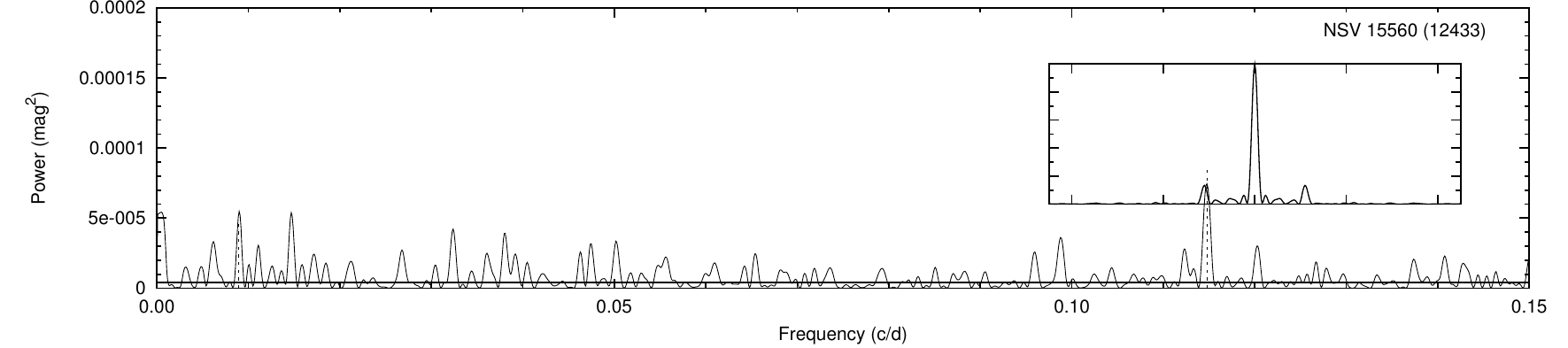}  % 012433
 \includegraphics[scale=1.0, angle=0]{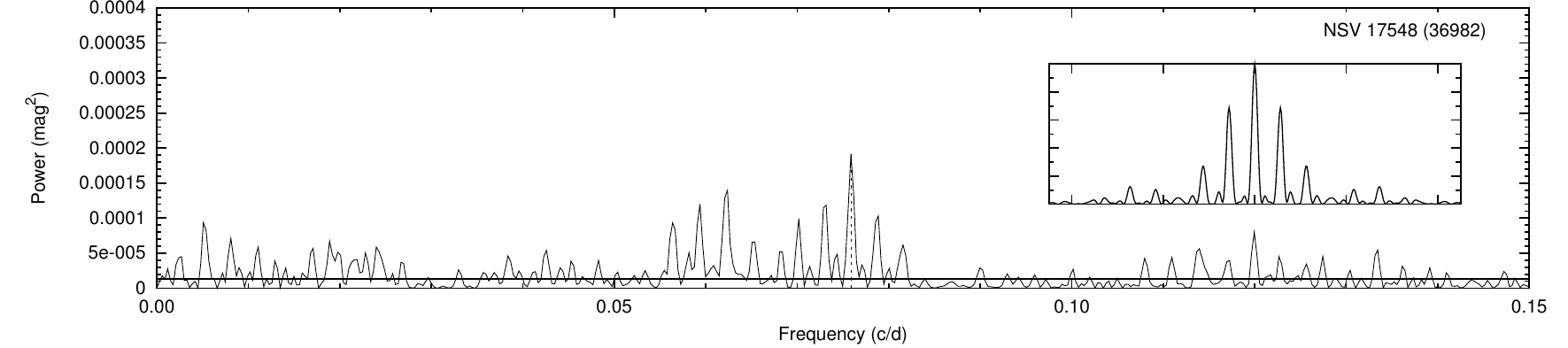}  % 017548
 \includegraphics[scale=1.0, angle=0]{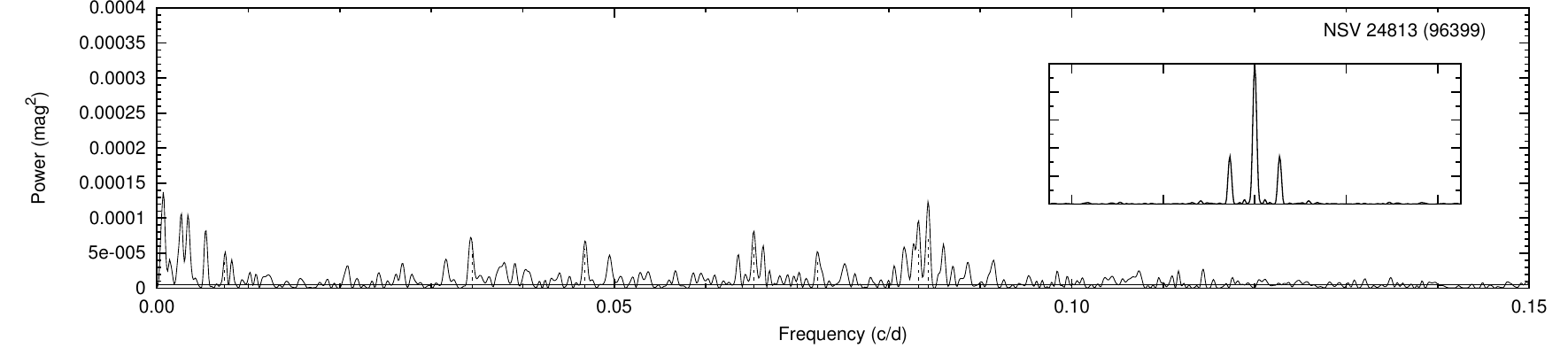}  % 096399
 \includegraphics[scale=1.0, angle=0]{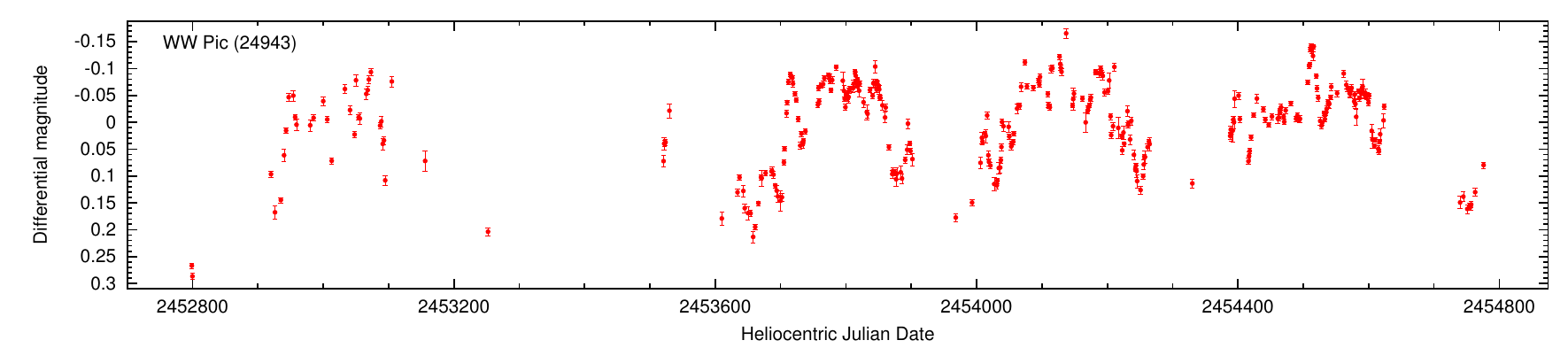}   % 024943
 \caption{Power spectra of selected stars exhibiting short-period pulsation (top 3 panels). The bottom panel is a light curve of WW Pic which exhibits a long secondary period of $\sim$ 370 d and a shorter 36-d period.}
 \label{fig023}
\end{figure*}

\bsp
\label{lastpage}
\end{document}